\documentclass [prb,twocolumn,amsmath,amssymb,longbibliography]{revtex4-2}
\setcitestyle{numbers,square}
\usepackage{appendix}
\usepackage{array}
\usepackage{amsmath,amssymb}
\usepackage{tabularx}
\usepackage{graphicx}% Include figure files \usepackage{dcolumn}% Align table columns on decimal point
\usepackage{bmpsize}
\usepackage{bm}% bold math
\usepackage{color}
\usepackage{amssymb}
\usepackage [version=3]{mhchem}
\usepackage{newtxmath}
\usepackage{url}
\usepackage{physics}
\usepackage{here}
\usepackage{float}
\usepackage{comment}
\usepackage{docmute}

\DeclareMathAlphabet{\mathpzc}{OT1}{pzc}{m}{it}

\newcommand{\rev} [1]{\textcolor{black}{ #1}}
\newcommand{\revblue} [1]{\textcolor{black}{ #1}}

\begin{document}
%\preprint{APS/123-QED}

%%%%%%%%%%%%%%%%%%%%%%%%%%%%%%%%%%%%%%%%%%%%%%%%%%%%%%%%%%%%%%%%%%%%%%%%%%%%%%%%%%%%%%%%%%%%%%%%%%%%%%%%%%%%%%%%%%%%%%%%%%%%%%%%
%%%%%%%%%%%%%%%%%%%%%%%%%%%%%%%%%%%%%%%%%%%%%%%%%%%%%%%%%%%%%%%%
\title{
{
Transfer Entropy and Flow of Information in Two-Skyrmion System}
}

\author{Tenta Tani$^1$}
\email{tenta.tani@qp.phys.sci.osaka-u.ac.jp}
\author{Soma Miki$^{2-5}$}
\email{soma.miki.d8@tohoku.ac.jp}
\author{Hiroki Mori$^2$}
\author{Minori Goto$^{2-4}$}
\author{Yoshishige Suzuki$^{2-4}$}
\author{Eiiti Tamura$^2$}

\affiliation{
\vspace{3mm}
$^1$Department of Physics, The University of Osaka, Toyonaka, Osaka 560-0043, Japan \\
$^2$Graduate School of Engineering Science, The University of Osaka, Toyonaka, Osaka 560-8531, Japan \\
$^3$Spintronics Research Network Division, Institute for Open and Transdisciplinary Research Initiatives, The University of Osaka, Suita, Osaka 565-0871, Japan \\
$^4$Center for Spintronics Research Network (CSRN), Graduate School of Engineering Science, The University of Osaka, Toyonaka, Osaka 560-8531, Japan \\
$^5$WPI Advanced Institute for Materials Research (AIMR), Tohoku University, Sendai, 980-8577, Japan
}

\date{\today}

%%%%%%%%%%%%%%%%%%%%%%%%%%%%%%%%%%%%%%%%%%%%%%%%%%%%%%%%%%%%%%%%%%%%%%%%%%%%%%%%%%%%%%%%%%%%%%%%%%%%%%%%%%%%%%%%%%%%%%%%%%%%%%%%
%%%%%%%%%%%%%%%%%%%%%%%%%%%%%%%%%%%%%%%%%%%%%%%%%%%%%%%%%%%%%%%%
\begin{abstract}
{
We theoretically investigate the flow of information in an interacting two-skyrmion system confined in a box at finite temperature. By numerical simulations based on the Thiele--Langevin equation, we demonstrate that the skyrmion motion cannot be fully described by the master equation, highlighting the nontrivial dynamics. Particularly, due to the chiral motion of skyrmion, we find asymmetric flow of information with violating the detailed balance condition. We analyze this system using information-theoretical quantities including Shannon entropy, mutual information, and transfer entropy. The physical significance of transfer entropy, which has been overlooked in previous studies, is elucidated. Notably, the peak position of the transfer entropy, as a function of time delay, is independent of the interaction range yet dependent on the box size. This peak corresponds to the characteristic time required for changing the skyrmion state. Due to the unusual asymmetric circulation of information, the two-skyrmion system can be a unique device for future applications to the natural computing.
}
\end{abstract}

\maketitle
%%%%%%%%%%%%%%%%%%%%%%%%%%%%%%%%%%%%%%%%%%%%%%%%%%%%%%%%%%%%%%%%%%%%%%%%%%%%%%%%%%%%%%%%%%%%%%%%%%%%%%%%%%%%%%%%%%%%%%%%%%%%%%%%
%%%%%%%%%%%%%%%%%%%%%%%%%%%%%%%%%%%%%%%%%%%%%%%%%%%%%%%%%%%%%%%%
\section{Introduction}
\label{sec:intro}

Information theory and physics are becoming increasingly interconnected, with one actively studied area being information thermodynamics~\cite{parrondo2015thermodynamics,andrieux2008nonequilibrium,jarzynski2008thermodynamics,hayashi2010fluctuation,sagawa2008second,sagawa2009minimal,sagawa2010generalized,sagawa2012fluctuation,Sagawa2013role,karbowski2024information}.
Information thermodynamics highlights the equivalence of information and energy (or work), as demonstrated by the resolution of Maxwell’s demon paradox~\cite{Szilard1929,brillouin1951maxwell,Bennett1982,landauer1961irreversibility,sagawa2008second,sagawa2009minimal,sagawa2010generalized,sagawa2012fluctuation,Sagawa2013role}.
Mutual information, an information-theoretical quantity, is associated with work, while transfer entropy~\cite{Schreiber2000}, a type of mutual information, serves as a measure of flow of information~\cite{Kobayashi2013,Kawasaki2014,Staniek2008,Vicente2011,Rozo2021,Kwon2008,marschinski2002,sandoval2014structure,falkowski2023causality,sun2014identification,gao2020single,murcio2015urban,ito2016backward,oka2013exploring,ito2013information}.

It is used to analyze the flow of information in various systems and time series, including neural systems~\cite{Kobayashi2013,Kawasaki2014,Staniek2008,Vicente2011} and stock markets~\cite{Kwon2008,marschinski2002}.
For example, Schreiber~\cite{Schreiber2000} introduced transfer entropy and used it to examine the flow of information between heartbeat and breathing, revealing a stronger flow of information from the heart rate to the breathing rate than that in the reverse direction, as indicated by its higher transfer entropy.

Although the transfer entropy has been mainly applied to complex systems and time series, its significance in simpler physical systems remains unclear.
For instance, in systems of interacting particles, the exact relationship between the transfer entropy and particle dynamics is unknown.
Key questions include: how can we define the flow of information between particles, and how is information transferred through particle collisions?
Addressing these questions in simpler systems can clarify the physical meaning of transfer entropy, enabling the application of flow of information concepts to various physical systems such as nanoparticles in liquids and molecules in gases. 
Ultimately, studying the transfer entropy in these systems is essential for advancing the interdisciplinary field connecting information theory and physics.

The skyrmion system~\cite{skyrme1962unified,roessler2006spontaneous,muhlbauer2009skyrmion,yu2011near} is ideal for studying flow of information.  Skyrmions exhibit Brownian motion~\cite{nozaki2019brownian,schutte2014inertia,zazvorka2019thermal,miki2021brownian,zhao2020topology,suzuki2021diffusion,jibiki2020skyrmion,miki2023spatial,ishikawa2021implementation,Weissenhofer2021skyrmion}, interact repulsively with one another~\cite{ishikawa2021implementation,lin2013particle,capic2020skyrmion}, and can be confined in an energetic box~\cite{jibiki2020skyrmion,miki2023spatial,ishikawa2021implementation}.
Stochastic Brownian motion enables information-theoretical analysis, while the repulsive interactions between skyrmions allow investigation of the flow of information.
The skyrmion system has already been experimentally realized in a confined box~\cite{jibiki2020skyrmion,miki2023spatial,ishikawa2021implementation}.
{In addition, skyrmion is promising for ultralow-power Brownian computing~\cite{Bennett1982,pinna2018skyrmion,zazvorka2019thermal} and an information entropy carrier~\cite{zivieri2022}.}
Therefore, understanding the nature of flow of information in skyrmion systems is valuable in fundamental and applied research.

In this study, we investigated a two-skyrmion system confined in a box at a finite temperature using the Thiel-Langevin equation as a theoretical model.
The analysis demonstrated that the simulated probability distribution cannot be reproduced by the master equation, indicating nontrivial skyrmion dynamics.
The chiral motion of skyrmion violates the detailed balance condition, meaning the asymmetric directional flow of information in equilibrium.
Therefore, the two-skyrmion system can be an interesting \rev{device} for more efficient machine learning algorithm using stochastic systems~\cite{Ohzeki2016stochastic,Takahashi2016,ohzeki2017quantum}.
Examination of the transfer entropy {$T_{Y \rightarrow X}(\Delta t)$} between ``the first skyrmion at time $t$'' and ``the second skyrmion at time $t - \Delta t$'', demonstrated that the functional shape of {$T_{Y \rightarrow X}(\Delta t)$} exhibits a peak structure.
The analysis also revealed that the peak position depends only on the box size but not on the skyrmion--skyrmion interaction range, suggesting that the transfer entropy peak corresponds to the time interval required for a skyrmion to influence another skyrmion through repulsive interactions---representing the information transmission time, \rev{as will be discussed in detail in Sec.~\ref{sec:entropic}.}
In fact, the estimated characteristic time, the box size divided by the average velocity of skyrmion, well match with the transfer entropy peak. 
Notably, we can understand that the information transmission time consists of the time to obtain mutual information and the time to write the information.
Owing to the simplicity of the system, these novel results on the nature of flow of information were revealed.
This is because previous research has focused on complex systems, where it is difficult to deeply analyze and understand the physical significance of transfer entropy.

This paper is organized as follows.
Sec.~\ref{sec:model} presents the theoretical model for examining the two-skyrmion system, introduces the Thiele--Langevin equation and discusses skyrmion interactions. Sec.~\ref{sec:dynamics} focuses on the skyrmion system dynamics, demonstrating their simplicity and nontrivial asymmetric flow of information.
Sec.~\ref{sec:entropic} explores several information-theoretical measures, including the transfer entropy to investigate the flow of information between two skyrmions.
Sec.~\ref{sec:conclusion} concludes the study.

%%%%%%%%%%%%%%%%%%%%%%%%%%%%%%%%%%%%%%%%%%%%%%%%%%%%%%%%%%%%%%%%%%%%%%%%%%%%%%%%%%%%%%%%%%%%%%%%%%%%%%%%%%%%%%%%%%%%%%%%%%%%%%%%
\section{Model of interacting skyrmions in a box}
\label{sec:model}

In this paper, we considered the classical motion of skyrmions. 
The dynamics of two interacting skyrmions in a potential box are governed by a stochastic differential equation called the Thiele--Langevin equation~\cite{thiele1973steady,schutte2014inertia,makhfudz2012inertia},
\begin{equation}
\begin{split}
    m \frac{d\bm{v}_1}{dt}
    &= -\alpha D \bm{v}_1 + \bm{G}\times\bm{v}_1 + \bm{F}_\mathrm{int}(\bm{r}_1, \bm{r}_2) + \bm{F}_\mathrm{wall}(\bm{r}_1) + \bm{R}_1,
    \\
    m \frac{d\bm{v}_2}{dt} 
    &= -\alpha D \bm{v}_2 + \bm{G}\times\bm{v}_2 - \bm{F}_\mathrm{int}(\bm{r}_1, \bm{r}_2) + \bm{F}_\mathrm{wall}(\bm{r}_2) + \bm{R}_2,
\end{split}
\label{eq:TL}
\end{equation}
where $\bm{r}_j = (x_j,y_j)$ and $\bm{v}_j=(v_{jx},v_{jy})$ are the two-dimensional position (the center of magnetization) and velocity of the $j$-th skyrmion $(j=1,2)$, respectively, and $m=1\times10^{-22} \, \mathrm{kg}$ is the skyrmion mass~\cite{schutte2014inertia,makhfudz2012inertia,buttner2015dynamics,suzuki2021diffusion,suzuki2022mass}.
In this study, mass is treated as a parameter.
%As presented later in Sec.~\ref{sec:entropic}, the relative values are important because the time constant depends on the mass; except at equilibrium, where each result is mass-independent.
%Although time constants of the investigated quantities such as transfer entropy and information flow depend on mass, this paper focuses on how results depend on parameters other than mass.
The present paper focuses on how investigated quantities, such as transfer entropy and information flow, depend on parameters other than mass.
This is because a change of mass value only alters the time constants of the results.
Thermal fluctuation-induced mass also exists~\cite{suzuki2021diffusion} in the system; however, it is negligible compared to the motion-induced mass in the potential~\cite{makhfudz2012inertia,buttner2015dynamics}. 
Therefore, only the latter is considered in this study.
\rev{We use the value of mass mentioned above since Ref.~\cite{buttner2015dynamics} experimentally reports $\sim 8 \times 10^{-22} \,\mathrm{kg}$.}
For the friction term $-\alpha D \bm{v}_j$, $\alpha=0.02$ is Gilbert constant, and $D$ is dissipation dyadic which depends on the skyrmion radius $R=25.55 \, \mathrm{nm}$ and domain wall width $w=11.35 \, \mathrm{nm}$~\cite{tamura2020skyrmion,cho2020manipulating,wang2018theory,hrabec2017current,belavin1975metastable}.
In the second term, $\bm{G} = -G\hat{\bm{z}}$ is the gyrocoupling vector~\cite{miki2021brownian,zhao2020topology,suzuki2021diffusion}.

\rev{The effective parameters $D$ and $G$ are derived from microscopic parameters as follows. 
They are expressed using the magnetization distribution $\bm{M}(\bm{r})$ inside the cylindrically symmetric skyrmion,
% \begin{equation}
%     \begin{split}
%         D &= \left( \frac{M_\mathrm{s}h}{|\gamma_e|} \right) \frac{1}{M_\mathrm{s}^2} \int \mathrm{d}^2\bm{r} \,\frac{\partial\bm{M}}{\partial x} \cdot \frac{\partial\bm{M}}{\partial x},
%         \\
%         G &= \left( \frac{4\pi M_\mathrm{s}h}{|\gamma_e|} \right) q,
%     \end{split}
% \end{equation}
\begin{align}
\label{eq:D-mag}
    D &= \left( \frac{M_\mathrm{s}h}{|\gamma_e|} \right) \frac{1}{M_\mathrm{s}^2} \int d^2\bm{r} \,\frac{\partial\bm{M}}{\partial x} \cdot \frac{\partial\bm{M}}{\partial x},
    \\
    G &= \left( \frac{4\pi M_\mathrm{s}h}{|\gamma_e|} \right) q,
\end{align}
where $M_\mathrm{s}$, $h$, and $\gamma_e = -1.76\times10^{11} \, \mathrm{T^{-1}s^{-1}}$ are the saturation magnetization, the film thickness, and the gyromagnetic ratio, respectively. $q$ is the skyrmion number defined as
\begin{equation}
    q = \frac{1}{4\pi M_\mathrm{s}^3} \int d^2\bm{r}\, \bm{M} \cdot \left( \frac{\partial\bm{M}}{\partial x} \times \frac{\partial\bm{M}}{\partial y} \right),
\end{equation}
which is an integer. 
In this paper, we consider the case of $q=+1$.
Here we use $M_\mathrm{s} = 580 \, \mathrm{kA/m}$ and $h = 1.2 \, \mathrm{nm}$~\cite{tamura2020skyrmion,cho2020manipulating,wang2018theory,hrabec2017current,belavin1975metastable}, leading to the numerical values $D=7.06\times10^{-14} \, \mathrm{kg/s}$ and $G = 4.97\times10^{-14} \, \mathrm{kg/s}$.
}
\revblue{To obtain the value of $D$, we integrate Eq.~\eqref{eq:D-mag} by using the skyrmion profile of a $360^\circ$ domain-wall model~\cite{wang2018theory} (see Supplemental Material S5~\footnote{(Supplemental material) The employed skyrmion profile and details of the calculation of dissipation dyadic is provided online (Sec. S5)}).}

We modeled the skyrmion interaction using an exponentially decaying interaction when the two skyrmions are apart, whereas a constant force upon contact expressed as
\begin{equation}
\bm{F}_\mathrm{int}(\bm{r}_1, \bm{r}_2) =
\begin{cases}
F_0 e^{-(r_{12}-R)/\xi} \displaystyle\frac{\bm{r}_{12}}{r_{12}} & (r_{12} > R) \\
F_0 \displaystyle\frac{\bm{r}_{12}}{r_{12}} & (r_{12} < R),
\end{cases}
\end{equation}
where $\bm{r}_{12} = \bm{r}_1 - \bm{r}_2$ and $r_{12} = |\bm{r}_{12}|$\rev{~\cite{capic2020skyrmion,lin2013particle}.}
$F_0 = 1\times10^{-11} \, \mathrm{N}$ is the magnitude and $\xi$ is the interaction range. 
For computational simplicity, this interaction was modeled as a short-range exchange interaction, neglecting the dipole interaction due to the small skyrmion radius of the order of $10^{-9}$ m. 
\rev{Note that the constant force for $r < R$ is a numerical regularization; however, in this study, we confirmed that the separation $r$ always satisfied $r > R$.}

The force exerted by the walls of the square box is modeled using exponentially decaying potential energy,
\begin{equation}
    \bm{F}_\mathrm{wall}(\bm{r}) = -\nabla U_\mathrm{wall}(\bm{r}),
\label{eq:Fwall}
\end{equation}
where the potential $U_\mathrm{wall}(\bm{r})$ is expressed as,
\begin{align}
\begin{split}
    U_\mathrm{wall}&(\bm{r}) =
    \\
    &U_0 \left[ e^{-(x+\tilde{d})/\xi_0} + e^{(x-\tilde{d})/\xi_0} + e^{-(y+\tilde{d})/\xi_0} + e^{(y-\tilde{d})/\xi_0} \right],
\end{split}
\label{eq:Uwall}
\end{align}
where $\tilde{d}$ is a parameter representing the size of the box.
In this equation, the magnitude $U_0$ was set to $U_0 = 1\times10^{-20} \, \mathrm{J}$ and $\xi_0 = 8.52 \, \mathrm{nm}$ to fit the results given by the analytical formula~\cite{tamura2020skyrmion,miki2021size} (see Supplemental Material S2~\footnote{(Supplemental material) The analytical formula for energetic box is provided online. (Sec. S2)}).
As shown in Fig.~\ref{fig:trajectory}, the side of the box is $2d$.
The box size $d=5.78R$ corresponds to the parameter $\tilde{d} = 3R$.

The last term $\bm{R}_j$ in Eq.~\eqref{eq:TL} is a random force originating from thermal fluctuations.
This force is determined through the fluctuation--dissipation theorem,
\begin{equation}
    [\bm{R}_j]_\alpha = \sqrt{\frac{2\alpha D k_\mathrm{B}T}{\delta t}} \mathcal{N}(0,1) \, (\alpha=x,y),
\end{equation}
where $\delta t$ is a time increment and $\mathcal{N}(0,1)$ stands for the stochastic variable with the standard normal distribution, $k_\mathrm{B}$ is the Boltzmann constant, and $T$ is the temperature of the system.
Hereafter, we set $T=300 \, \mathrm{K}$.

{
Experimentally, a technique to confine two skyrmions in a box~\cite{ishikawa2021implementation} have been reported.
Furthermore, an experimental analysis of the confined two-skyrmion system in terms of transfer entropy have been also reported~\cite{suzuki2025informationdynamicsnaturalcomputing}.
}

%%%%%%%%%%%%%%%%%%%%%%%%%%%%%%%%%%%%%%%%%%%%%%%%%%%%%%%%%%%%%%%%%%%%%%%%%%%%%%%%%%%%%%%%%%%%%%%%%%%%%%%%%%%%%%%%%%%%%%%%%%%%%%%%
%%%%%%%%%%%%%%%%%%%%%%%%%%%%%%%%%%%%%%%%%%%%%%%%%%%%%%%%%%%%%%%%
\section{Skyrmion dynamics}
\label{sec:dynamics}

\begin{figure}
\begin{center}
   \includegraphics [width=0.8\linewidth]{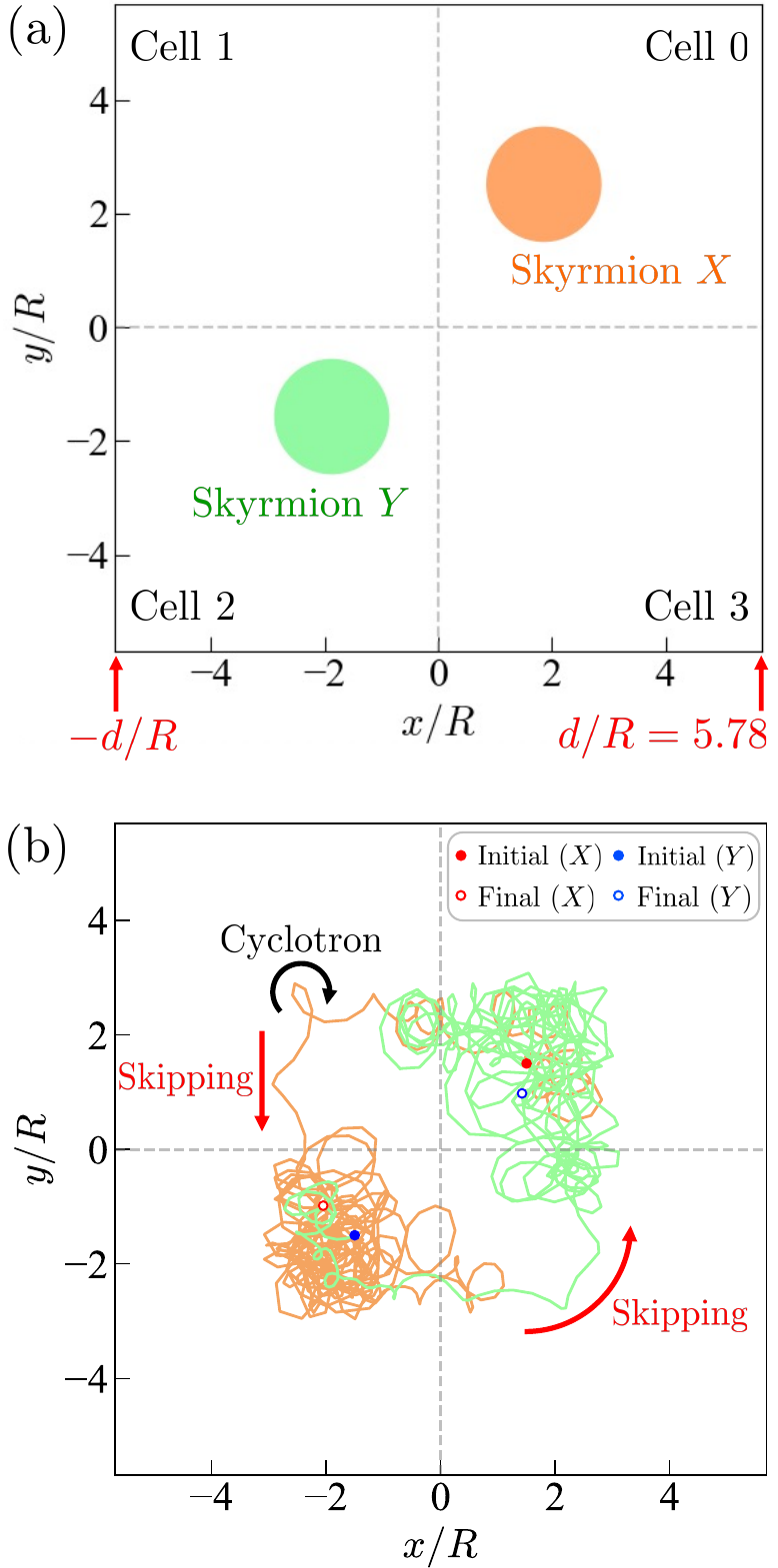}
   \caption{
   (Color online) (a)~Snapshot of the two skyrmions drawn by the two circles with radius $R$, in a square box.
   The box is divided into the four cells (cell 0, 1, 2, and 3).
   (b)~Trajectories of the two skyrmions.
   The initial and final points are shown by the filled and open circles, respectively.
   We see the clockwise cyclotron motion and the counterclockwise skipping trajectory.
            }\label{fig:trajectory}
 \end{center}
 \end{figure}

By solving the Thiele--Langevin equation [Eq.~\eqref{eq:TL}], we have derived the trajectories of the two skyrmions.
Fig.~\ref{fig:trajectory}(a) illustrates a snapshot of their motions, where the first and the second skyrmions are labeled as ``skyrmion $X$'' and ``skyrmion $Y$'', respectively.
The box size was set to $d=5.78R \, (\tilde{d}=3R)$.
To information-theoretically analyze the dynamics of skyrmions in Sec.~\ref{sec:entropic}, the skyrmions positions were discretized into four cells: cells 0–3 [see Fig.~\ref{fig:trajectory}(a)].
\rev{This four-cell discretization of a particle's continuous position has been theoretically and experimentally considered in the context of the cellular automata~\cite{ishikawa2021implementation,Lent1993,Lent1994,Snider1998,Amlani1999,Bahar2015,arrighi2019overview,Farrelly2020review}.}
Fig.~\ref{fig:trajectory}(b) illustrates the trajectories of the two skyrmions, illustrating confinement within the box and mutual repulsion.
{The clockwise cyclotron motion caused by the gyrocoupling term $\bm{G}\times\bm{v}$ [see Eq.~\eqref{eq:TL}] is analogous to the Lorentz force on an electron in a magnetic field.
In addition, one finds the counterclockwise skipping motion along the walls.}

We based the analysis on $N \approx 10^5$ simulations, which are sufficient to minimize statistical errors.
Throughout this study, we used a time increment $\delta t=1 \, \mathrm{ns}$ and a simulation time $t_\mathrm{f} = 1000 \, \mathrm{ns}$.
We employed the fourth-order Runge--Kutta method\rev{, which corresponds to the Stratonovich calculus.}
\rev{
Note that since the coefficient of the random force $\sqrt{2\alpha Dk_\mathrm{B}T / \delta t}$ is constant (i.e., additive noise), the It\^{o} and Stratonovich calculus yield the same results in this system.
The method for the simulation is explained in detail in the Supplemental Material S4~\footnote{(Supplemental material) The computational detail of the simulation is provided online (Sec. S4)}.
}

In Sec.~\ref{subsec:onebody}, we consider a single skyrmion system, followed by an interacting two-skyrmion system in Sec.~\ref{subsec:twobody}.

%%%%%%%%%%%%%%%%%%%%%%%%%%%%%%%%%%%%%%%%%%%%%%%%%%%%%%%%%%%%%%%%%%%%%%
\subsection{One-body system}
\label{subsec:onebody}

First, we consider a single skyrmion in the box.
\rev{By simulating the skyrmion's motion in the continuous space, we obtain the continuous trajectory $\bm{r}_1(t) = (x_1(t),y_1(t))$. Then, we transform this to the discretized value $x_t$.}
We denote the discretized position of the skyrmion at time $t$ in $i$-th simulation as $x_t^{(i)} = 0,1,2,3$.
Numerical calculations yield $N$ time series
\begin{equation}
    (x_0^{(i)}, x_{\delta t}^{(i)}, x_{2\delta t}^{(i)}, ..., x_{t_\mathrm{f}}^{(i)}), \quad i=1,2, ..., N.
\end{equation}
At each time $t$, the state of the skyrmion was treated as a stochastic variable $X_t$.
The probability that a skyrmion exists in cell $x_t$ at time $t$ is given by
\begin{equation}
    p(x_t) = \frac{1}{N}\sum_{i=1}^{N} \delta(x_{t}^{(i)} - x_{t}),
\end{equation}
where $\delta(x)$ is Kronecker's delta,
\begin{equation}
\delta(x) =
\begin{cases}
1 & (x=0) \\
0 & (x \neq 0).
\end{cases}
\end{equation}
Similarly, the joint probability between different times $t$ and $t'$ is written as
\begin{equation}
    p(x_t, x_{t'}) = \frac{1}{N}\sum_{i=1}^{N} \delta(x_{t}^{(i)} - x_{t}) \delta(x_{t'}^{(i)} - x_{t'}),
\end{equation}
and the conditional probability can be obtained by Bayes' theorem: $p(x_t \mid x_{t'}) = p(x_t, x_{t'}) / p(x_{t'})$.
\begin{figure}
\begin{center}
   \includegraphics [width=0.85\linewidth]{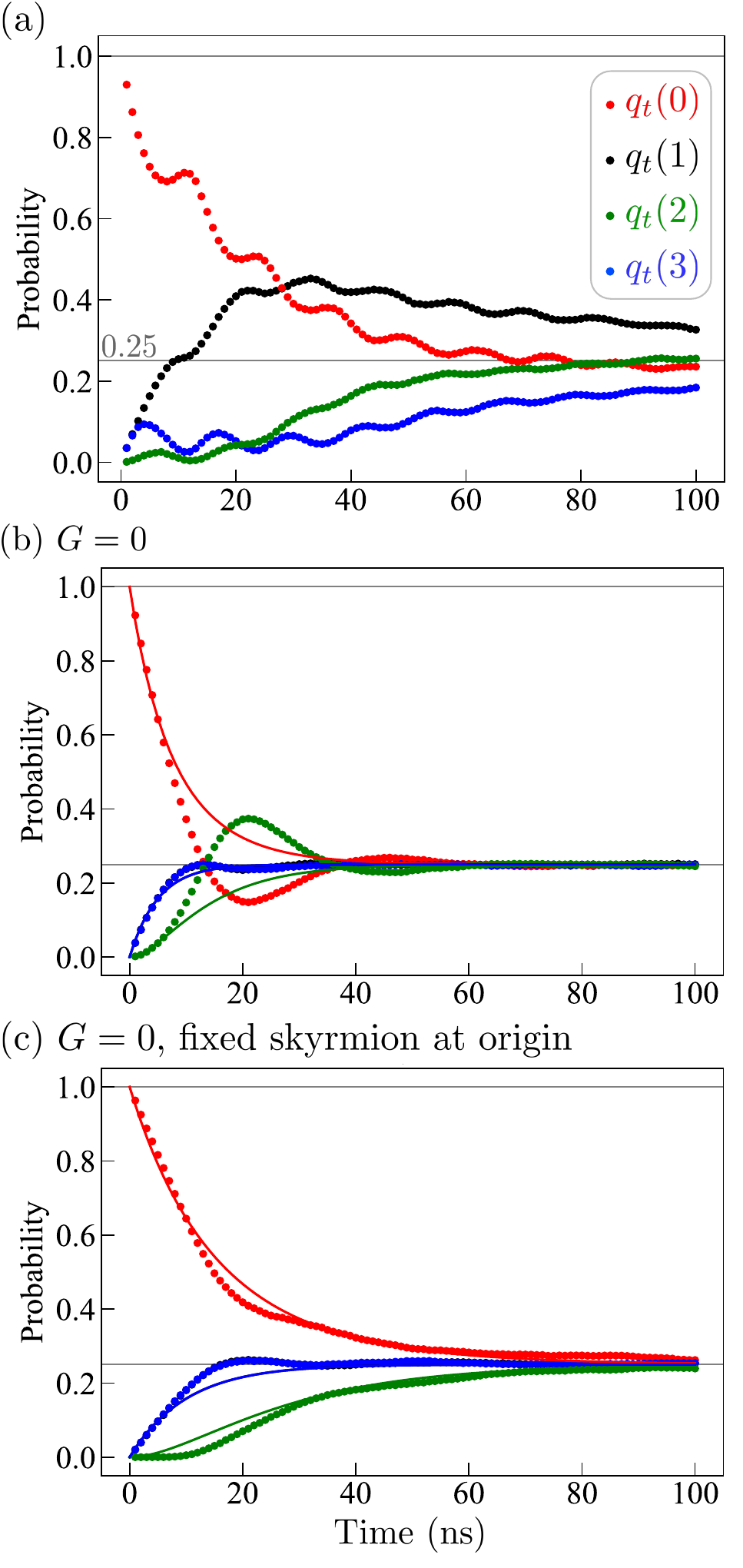}
   \caption{
   (Color online) Time evolution of the occupation probabilities $q_t(x)$ for $x=0,1,2,3$ for (a)~single skyrmion, (b)~single skyrmion with $G=0$, and (c)~single moving skyrmion and fixed skyrmion at the origin, with $G=0$.
   The dots are the simulated results, and the solid curves are the analytical solutions of the master equation.
   The horizontal line indicates $q=0.25$, to which all probabilities converge in $t \rightarrow \infty$ limit.
            }\label{fig:probability}
 \end{center}
 \end{figure}
Based on this formulation, we investigated the time evolution of the occupation probability.
In the simulation, we set the initial position $\bm{r}_1(t=0) = (\tilde{d}/2, \tilde{d}/2) = (1.5R, 1.5R)$ and the initial velocity $\bm{v}_1(t=0) = (0,0)$.

\rev{
Figure~\ref{fig:probability}(a) illustrates the conditional probability $q_t(x) = p(x_{t_\mathrm{f}} = x \mid x_{t_\mathrm{f} - t} = 0)$, where we observe an equilibrium state [$t_\mathrm{f}-t > 200 \, \mathrm{ns}$; see Sec.~\ref{sec:entropic} and the non-interacting case in Fig.~\ref{fig:ShannonMI}(a)].
}
Since initial condition is set to $x = 0$, $q_t(0)$ rapidly decreases from $1$ to $0.25$. 
The wavy probability profiles result from the clockwise cyclotron motions induced by the gyrocoupling term $\bm{G}\times\bm{v}$, with the oscillation frequency matching the cyclotron frequency $\omega = G/m$.
The asymmetry between $q_t(1)$ and $q_t(3)$ is attributed to the skipping trajectory along the walls.
This skipping is simply an edge current accompanied by cyclotron motion; thus, its direction is counterclockwise.
These clockwise and counterclockwise motions cancel out in the sense of angular momentum, obeying Bohr--van Leeuwen theorem.
However, the clockwise cyclotron motion is not captured because of the coarse-graining using the four cells.
{As a result, the Bohr--van Leeuwen theorem appears to be violated. The system virtually has a finite angular momentum.
\rev{This fictitious angular momentum was experimentally observed~\cite{miki2021brownian}; due to a low video frame rate, only the global skipping trajectory and corresponding finite angular momentum was observed.}
The sensitivity of our results on the number of cells is discussed in Sec.~\ref{sec:entropic}.
}

{Due to the chiral motion of the skyrmion, the detailed balance condition also appears to be violated.}
% However, the clockwise cyclotron motion is not captured because of the coarse-graining using the four cells.
% As a result, the detailed balance theorem appears to be violated.
\begin{figure}
\begin{center}
   \includegraphics [width=0.85\linewidth]{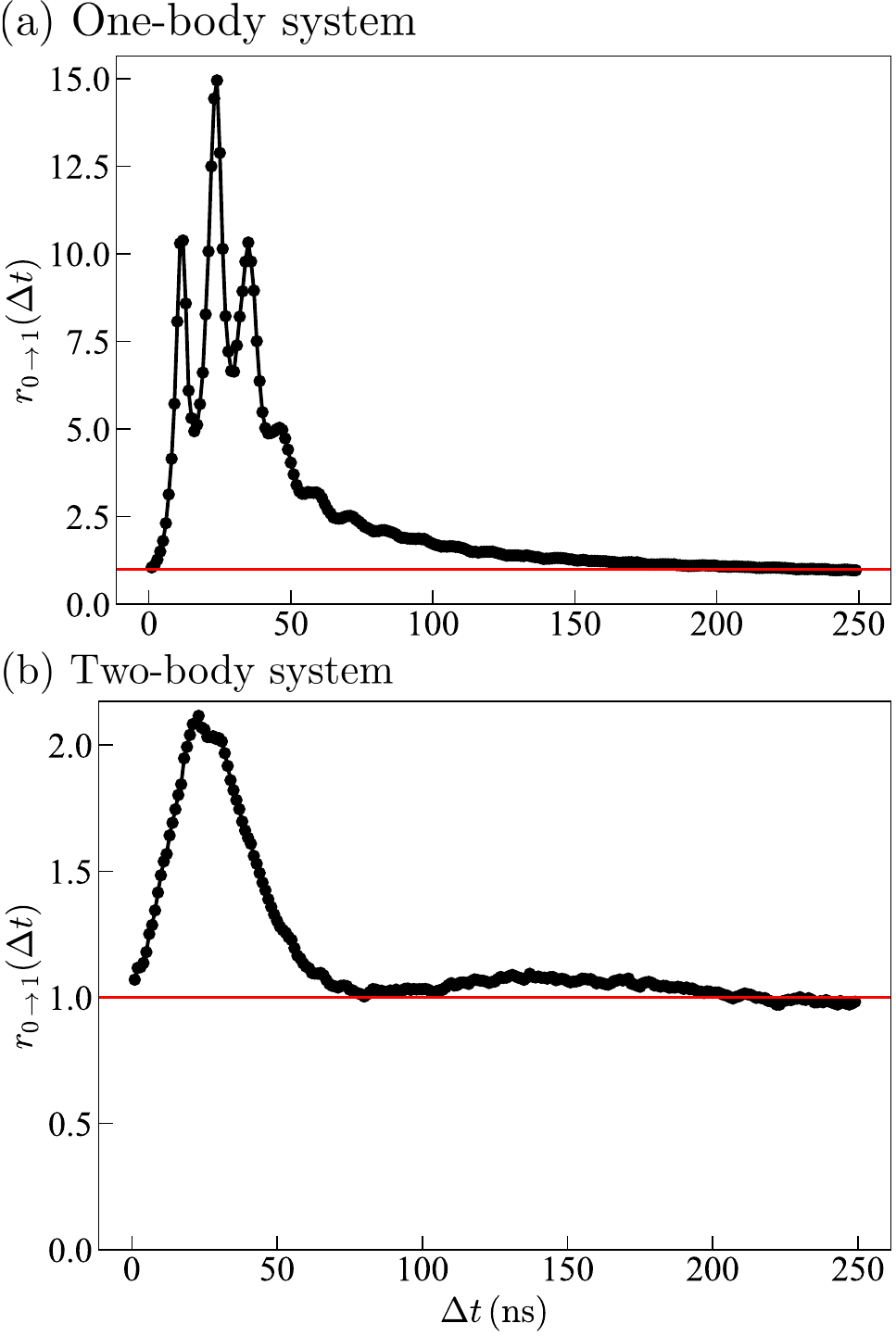}
   \caption{
   (Color online) Probability ratio $r_{0\rightarrow1}(\Delta t)$ as a function of the time delay $\Delta t$.
   (a) and (b) is for the one-body and two-body cases, respectively.
   The horizontal lines indicate $r_{0\rightarrow1} = 1$, or the detailed balance condition.
            }\label{fig:detail}
 \end{center}
 \end{figure}
{In Fig.~\ref{fig:detail}(a), we show the ratio between the forward ($0 \rightarrow 1$) and backward ($1 \rightarrow 0$) process,}
\begin{equation}
    r_{0\rightarrow1} (\Delta t) =
    \frac{p(x_{t-\Delta t}=0) p(x_t=1 \mid x_{t-\Delta t}=0)}
    {p(x_{t-\Delta t}=1) p(x_t=0 \mid x_{t-\Delta t}=1)}.
\label{eq:detail}
\end{equation}
It should be noticed that in the above equation $t=t_\mathrm{f}$ was employed.
As we will see in Sec.~\ref{sec:entropic}, the system is in equilibrium at $t=t_\mathrm{f}$.
The detailed balance condition $r_{0\rightarrow1} = 1$ (the horizontal lines in Fig.~\ref{fig:detail}) is clearly violated.
The ratio $r_{0\rightarrow1}$ exceeds unity and converges into $r_{0\rightarrow1} = 1$ with increasing $\Delta t$.
Therefore, information circulates in the counterclockwise direction.
{We note that the definition [Eq.~\eqref{eq:detail}] does not include the sign reversal of $G$, the direction of the skyrmion core.
If we flip the sign of $G$ in the backward process, the detailed balance condition is recovered.
However, in applications using the skyrmion dynamics, the magnetization is not reversed, and therefore, the apparent violation of the detailed balance is realized. 
}

When $G$ is turned off, probabilities evolve as illustrated in Fig.~\ref{fig:probability}(b).
Without chiral motion, $q_t(1) = q_t(3)$ at all times.
Notably, near $t=20 \, \mathrm{ns}$, $q_t(2)$ temporarily exceeds $q_t(0)$ which is a nontrivial overshoot behavior that cannot be reproduced by the results obtained using the master equation [solid curves in Fig.~\ref{fig:probability}(b)].
For the curve fitting, we used the transition rate parameters $w=w'=0.05$.
A detailed description of the master equation and its analytical solution is provided in Supplemental Material S1~\cite{supplemastereq}.
The nontrivial overshooting behavior in Fig.~\ref{fig:probability}(b) is caused by the singularity of the center of the box.
As illustrated in Fig.~\ref{fig:probability}(c), fixing the skyrmion at the origin eliminates the overshoot, allowing the master equation to qualitatively match the dynamics with $w=w'=0.025$ (see Ref.~\cite{supplemastereq}).
Moreover, if the square box is divided into two cells, the master equation reproduces the simulated probability distribution very well (data not shown); however, overshooting persists without a skyrmion fixed at the origin.

%%%%%%%%%%%%%%%%%%%%%%%%%%%%%%%%%%%%%%%%%%%%%%%%%%%%%%%%%%%%%%%%%%%%%%
\subsection{Two-body system}
\label{subsec:twobody}

\begin{figure}
\begin{center}
   \includegraphics [width=0.85\linewidth]{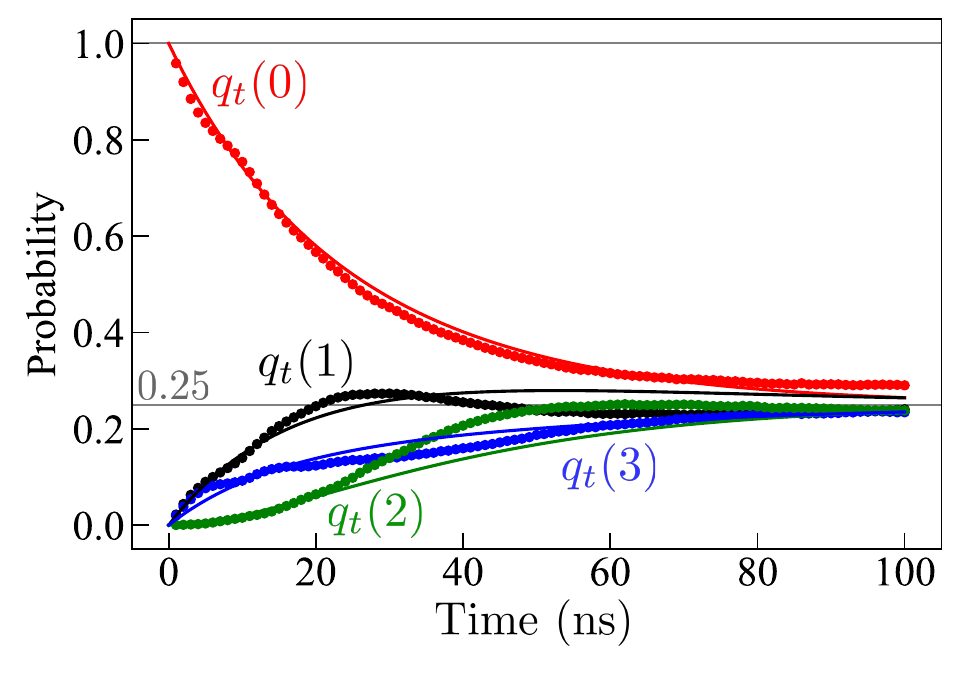}
   \caption{
   (Color online) For the two-skyrmion system, time evolution of the occupation probabilities $q_t(x)$ for $x=0,1,2,3$ are shown.
   The dots are the result of the simulation, and the solid lines are given from the analytical solutions of the master equation ($w=0.02$ and $w'=0.012$).
            }\label{fig:prob2body}
 \end{center}
 \end{figure}

For the rest of this study, we consider two interacting skyrmions confined in a box.
The simulation is based on the same condition as Sec.~\ref{subsec:onebody}.
The initial states of the skyrmion $X$ and $Y$ are set to be $\bm{r}_1(t=0) = (\tilde{d}/2, \tilde{d}/2)$ and $\bm{r}_2(t=0) = (-\tilde{d}/2, -\tilde{d}/2)$, respectively.
\rev{Here, the continuous trajectory $\bm{r}_1(t)$ and $\bm{r}_2(t)$ are transformed to the discretized values $x_t$ and $y_t$, respectively.}
The both of the skyrmions have no initial velocities.
The time evolution of the occupation probability $q_t(x)$ of the skyrmion $X$ is shown in Fig.~\ref{fig:prob2body}.
The skyrmion--skyrmion interactions suppress the oscillatory behavior in the single-skyrmion case.
The master equation curve fit was also unsuccessful in capturing the behavior of the system (solid curves in Fig.~\ref{fig:prob2body}), with parameters $w=0.02$ and $w'=0.012$ (see Ref.~\cite{supplemastereq}).

The detailed balance condition is violated also in the two-skyrmion case.
In Fig.~\ref{fig:detail}(b), we illustrate the probability ratio $r_{0\rightarrow1} (\Delta t)$ [Eq.~\eqref{eq:detail}].
The values of $r_{0\rightarrow1}$ is suppressed by clockwise motions due to the skyrmion--skyrmion interaction.
However, an asymmetric flow of information in the counterclockwise direction still exists.

\rev{It is known that the introduction of forces violating the detailed balance condition accelerates the convergence to the equilibrium state in stochastic gradient Langevin dynamics~\cite{Ohzeki2016stochastic}. 
The gyrotropic force of skyrmions inherently plays this role, suggesting a potential physical implementation of such accelerated sampling algorithms.}
%Since the violation of the detailed balance condition was shown to realize more efficient machine learning algorithm~\cite{Ohzeki2016stochastic,Takahashi2016,ohzeki2017quantum}, the two-skyrmion system can be an intriguing platform for machine learning using stochastic systems.
%Specifically, the violation of the detailed balance condition is proposed to enhance the efficiency of the stochastic gradient descent method, which is for the large data set for the learning.
We note that the detailed balance condition is already violated at the level of the master equation, if the transition rate is asymmetric, i.e., $w \neq w'$ (see Ref.~\cite{supplemastereq}).

{This system is simple in the sense that there is the equation of motion [Eq.~\eqref{eq:TL}], which does not exist in the complex systems treated in the previous research on transfer entropy, such as neural systems~\cite{Kobayashi2013,Kawasaki2014,Staniek2008,Vicente2011} and stock markets~\cite{Kwon2008,marschinski2002}.
Given that, it is worth noting that the two-skyrmion system has the rich dynamics, including violating the detailed balance condition.
}

%%%%%%%%%%%%%%%%%%%%%%%%%%%%%%%%%%%%%%%%%%%%%%%%%%%%%%%%%%%%%%%%%%%%%%%%%%%%%%%%%%%%%%%%%%%%%%%%%%%%%%%%%%%%%%%%%%%%%%%%%%%%%%%%
\section{Information-theoretical analysis using entropic quantities}
\label{sec:entropic}

In this section, we analyze the motions of the two skyrmions using entropic quantities to investigate the nature of the transfer entropy.
Let $X_t, Y_t$ be the stochastic variables of the skyrmion $X$ and $Y$ at time $t$, respectively.
The Shannon entropy of skyrmion $X$ at time $t$ is written as
\begin{equation}
    H(X_t) = -\sum_{x_t} p(x_t) \ln{p(x_t)},
\end{equation}
which quantifies the randomness of the position of the skyrmion $X$.
Similarly, we can calculate the conditional Shannon entropy between $X_t$ and $Y_{t-\Delta t}$ using
\begin{equation}
    H(X_t \mid Y_{t-\Delta t}) = -\sum_{x_t, y_{t-\Delta t}} p(x_t, y_{t-\Delta t}) \ln{p(x_t \mid y_{t-\Delta t})}.
\end{equation}
\begin{figure}
\begin{center}
   \includegraphics [width=0.85\linewidth]{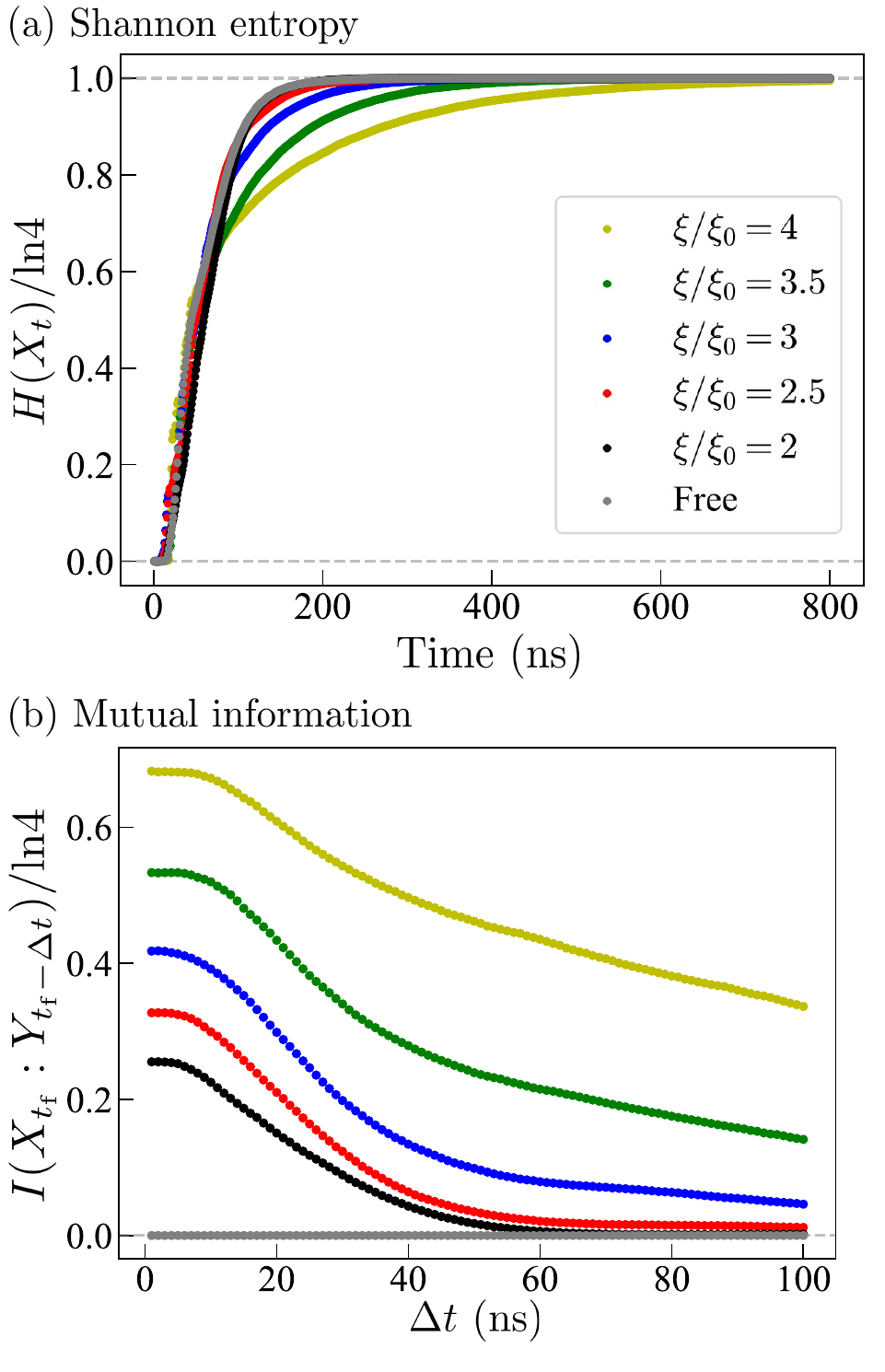}
   \caption{
   (Color online) (a)~Shannon entropy $H(X_t)$ of the skyrmion in units of $\ln{4}$ for several interaction ranges $\xi/\xi_0 = 2, 2.5, 3, 3.5, 4$.
   We show the non-interacting case (free skyrmion) as well.
   (b)~Mutual information $I(X_t:Y_{t-\Delta t})$ is plotted as a function of the time delay $\Delta t$, in the same manner as (a).
            }\label{fig:ShannonMI}
 \end{center}
 \end{figure}
Fig.~\ref{fig:ShannonMI}(a) illustrates the time evolution of Shannon entropy $H(X_t)$ in units of $\ln{4}$ for various interaction ranges $\xi$.
In addition to the results of $\xi/\xi_0 = 2, 2.5, 3, 3.5, 4$, we present the case of a free skyrmion (no interaction).
The box size is fixed at $d/R=5.78$.
Due to symmetry, $H(X_t) = H(Y_t)$.
In all cases, the Shannon entropy converges to the equilibrium state ($H=\ln{4}$) for $t > 800 \, \mathrm{ns}$.
The skyrmion--skyrmion interaction delays relaxation to the equilibrium state, thus cases with longer interaction range $\xi$ are slower in reaching equilibrium.

To quantify the correlation between the two skyrmions, we used the mutual information
\begin{equation}
    I(X_t : Y_{t-\Delta t}) = H(X_t) - H(X_t \mid Y_{t-\Delta t}),
\end{equation}
which consists of the Shannon entropy and conditional Shannon entropy.
Mutual information $I(X_t : Y_{t-\Delta t})$ represents the information shared between skyrmion $X$ at time $t$ ($X_t$) and skyrmion $Y$ at time $t-\Delta t$ ($ Y_{t-\Delta t}$).
At large time delays $\Delta t$, $X_t$ and $Y_{t-\Delta t}$ become uncorrelated such that $H(X_t \mid Y_{t-\Delta t}) = H(X_t)$ and $I(X_t : Y_{t-\Delta t}) = 0$, indicating no shared information.
In Fig.~\ref{fig:ShannonMI}(b), $I(X_t : Y_{t-\Delta t})$ is illustrated as a function of the time delay $\Delta t$ under the same condition as in Fig.~\ref{fig:ShannonMI}(a).
To consider the equilibrium states, we set $t=t_\mathrm{f}=1000 \, \mathrm{ns}$.
%For all interaction ranges $\xi$, the mutual information decreses monotonically with increasing $\Delta t$, as the equal-time correlation at $\Delta t = 0$ diminishes due to Brownian motion.
The mutual information can be easily lost, since only a random movement of one of the pair affects the shared information.
Therefore, the equal-time mutual information at $\Delta t = 0$, namely $I(X_{t_\mathrm{f}} : Y_{t_\mathrm{f}})$, is only kept in the time interval $< 10 \, \mathrm{ns}$.
We note that this time range, or the shoulder part in the plot, becomes longer with increasing $\xi$.
Stronger interactions (larger $\xi$) result in greater mutual information, while free skyrmions demonstrate no mutual information.
%Thus, these results confirm that $I(X_t : Y_{t-\Delta t})$ does not represent the flow of information between two skyrmions.
As it was pointed by Horowitz et al.~\cite{horowitz2014}, $(I(X_t : Y_{t-\Delta t}) - I(X_t : Y_t)) / \Delta t$ expresses an internal information flow, which realize an observation and a feedback process in the system (Maxwell's demon's task).
Here, the initial slope of the graph is exactly zero showing the absence of the Maxwell's demon in the simple two-skyrmion system.

\begin{figure*}
\begin{center}
   \includegraphics [width=\linewidth]{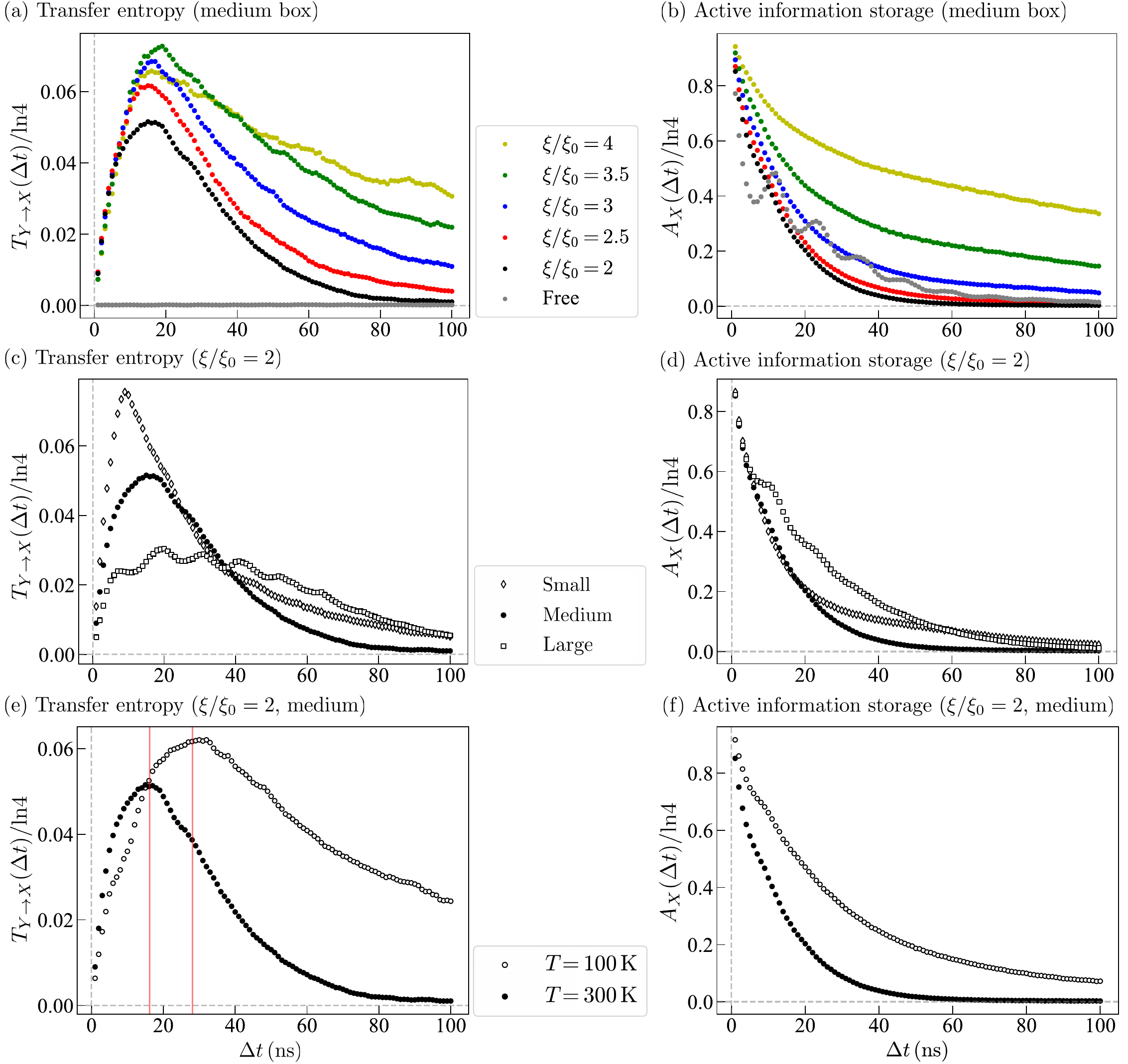}
   \caption{ 
   (Color online) For various interaction ranges $\xi/\xi_0 = 2, 2.5, 3, 3.5, 4$, (a)~the transfer entropy and (b)~the active information storage are plotted as a function of the time delay $\Delta t$, where the box size $d/R=5.78$ (medium size).
   %The meaning of the different colors is the same as Fig.~\ref{fig:ShannonMI}.
   (c) and (d) are the same plots for the three different box sizes.
   The respective small, medium, and large boxes correspond to rhombus, circle, and square dots.
   The interaction range $\xi/\xi_0=2$.
   In (e) and (f), the temperature is changed: $T = 100~\mathrm{K}$ and $300~\mathrm{K}$, with the interaction range $\xi/\xi_0=2$ and the medium box size $d/R=5.78$.
   The vertical lines indicate $d/v$ for each temperature (see the text).
            }\label{fig:TE}
 \end{center}
 \end{figure*}

To explore the directional flow of information, we investigated the transfer entropy
\begin{align}
\begin{split}
    T_{Y \rightarrow X}(\Delta t) &= I(X_t : Y_{t-\Delta t} \mid X_{t-\Delta t})
    \\
    &= H(X_t \mid X_{t-\Delta t}) - H(X_t \mid X_{t-\Delta t}, Y_{t-\Delta t}),
\label{eq:TE}
\end{split}
\end{align}
which is a widely used measure of flow of information~\cite{Kobayashi2013,Kawasaki2014,Staniek2008,Vicente2011,Rozo2021,Kwon2008,marschinski2002,sandoval2014structure,falkowski2023causality,sun2014identification,gao2020single,murcio2015urban,ito2016backward,oka2013exploring,ito2013information}.
Fig.~\ref{fig:TE}(a) shows the calculated transfer entropy $T_{Y \rightarrow X}(\Delta t)$ as a function of the time delay $\Delta t$ for various interaction ranges $\xi/\xi_0$ with $t$ fixed at $t_{\mathrm{f}}=1000 \, \mathrm{ns}$ and $d/R = 5.78$.
Unlike the mutual information illustrated in Fig.~\ref{fig:ShannonMI}(b), the transfer entropy exhibits a pronounced peak structure.
Notably, the peak positions remain nearly unchanged across all interaction ranges $\xi$.
As $\xi$ increases, the peak magnitude of the transfer entropy generally grows, except for $\xi/\xi_0 = 4$.
In the extreme case of $\xi/\xi_0 = 4$, the two skyrmions become less active.
Due to the extended interaction range, the skyrmions experienced strong repulsion even when they were far apart and located near a corner.
Consequently, the skyrmions tend to reside in stable cells (e.g., cell 0 and 2), thereby suppressing the peak value of the transfer entropy.
In addition to transfer entropy, we consider active information storage~\cite{LIZIER201239},
\begin{equation}
    A_X (\Delta t) = I(X_t : X_{t-\Delta t}),
\end{equation}
which can be regarded as the transfer entropy between the skyrmion $X$ and itself.
Fig.~\ref{fig:TE}(b) shows $A_X(\Delta t)$ at $t=t_\mathrm{f}$ as a function of $\Delta t$.
Unlike the previous form of transfer entropy, $A_X(\Delta t)$ exhibits no peak structures, indicating a monotonous memory loss due to the random motion of the skyrmion.
However, the forgetting speed is slower for longer interaction ranges $\xi$, because stronger interactions suppress random cell transitions.
Notably, the wavy structure of the interaction-free skyrmion arises from the cyclotron motion (see Sec.~\ref{sec:dynamics}), which vanishes in interacting cases.
Together, Figs.~\ref{fig:TE}(a) and (b) reveal that the transfer entropy peak originates purely from the skyrmion--skyrmion interactions.

Moreover, the impact of box size is examined for $\xi/\xi_0 = 2$, as illustrated in Figs.~\ref{fig:TE}(c) and (d) which present the results for transfer entropy and active information storage, respectively.
Small ($\tilde{d}=2R$), medium ($\tilde{d}=3R$), and large ($\tilde{d}=4R$) boxes are represented by the rhombus, circle, and square dots, respectively.
In Fig.~\ref{fig:TE}(c), unlike in Fig.~\ref{fig:TE}(a), the peak positions of $T_{Y \rightarrow X}(\Delta t)$ shift depending on the box size $d$; larger box sizes $d$ lead to later peak time.
For the large box, cyclotron motion becomes evident as the two skyrmions are free to move at the corners of the box.
In Fig.~\ref{fig:TE}(d), the active information storage continues to monotonously decay, similar to Fig.~\ref{fig:TE}(b).
Compared with the case of the medium-size box (filled circles), the skyrmions in the smaller box (rhombus dots) lose their memory more slowly.
Skyrmions confined in the smaller box interact more strongly each other and tend to remain in the same stable configuration (e.g. cell 0 and 2).
In the large box (square dots), the forgetting speed is slower than the medium-size box (filled circles), because the two skyrmions tend to move freely near the corners.

Comparing the mutual information [Fig.~\ref{fig:ShannonMI}(b)] and the transfer entropy [Fig.~\ref{fig:TE}(a)] for several $\xi$, we find that the mutual information starts to decrease at certain time, which is earlier than the transfer entropy peak.
This is because mutual information represents the shared information between skyrmion $X$ and $Y$: whichever skyrmion randomly moves, the mutual information drops.
Additionally, note that the shoulder of the mutual information [Fig.~\ref{fig:ShannonMI}(b)] becomes longer as $\xi$ increases, demonstrating that stronger interaction contributes to keep the shared information between the skyrmions.

From these results, the transfer entropy peak position can be interpreted as the time required for state transitions driven by skyrmion--skyrmion interaction.
For example, if skyrmion $X$ is located in cell $0$ and skyrmion $Y$ in cell $2$ (a stable state), a random force might accidentally move $Y$ to cell $3$, prompting $X$ to shift to cell $1$ due to repulsive interactions.
However, these state transitions ($Y:2\rightarrow3$ and $X:0\rightarrow1$) do not occur instantaneously but occur in finite time, which depends on the box size $d$.
Smaller boxes enable faster state transitions, resulting in earlier transfer entropy peaks.
This interval represents the information transmission time necessary for one skyrmion to influence the other.
Since the peak position is unchanged when we vary interaction ranges $\xi$, we can understand information transmission time consists of the time to obtain mutual information and the time to write the information.
In our two-skyrmion system, mutual information exists from the beginning, therefore the information transmission time is equal to the writing time of information.
%In fact, the transfer entropy peak positions, which depend on the box size $d$, are qualitatively reproduced using the transition time between adjacent cells by diffusion.

We can confirm this consequence by evaluating a characteristic time $d/v$, where $v = \sqrt{2k_\mathrm{B} T/m}$ is the average velocity of the skyrmions.
Although the skyrmions in the system interact with each other, we can safely use this velocity formula.
We confirmed that, by the interaction, the skyrmions do not move in the radial direction but move in the tangential direction.
\rev{The observed suppression of radial motion is an intrinsic consequence of the gyrocoupling term $\bm{G} \times \bm{v}$ in the Thiele equation. Because this term directs the velocity perpendicular to the force, the repulsive interaction between skyrmions results primarily in tangential circular motion rather than radial motion. We confirmed this mechanism by performing control simulations with $G=0$, which resulted in significant radial motion.}
Therefore, the absolute value of the velocity is almost unchanged by the skyrmion--skyrmion interaction.

In Fig.~\ref{fig:TE}(e) and (f), we show the transfer entropy and the active information storage for different temperatures 100 K (open circles) and 300 K (filled circles), respectively.
The interaction range $\xi/\xi_0 = 2$ and the box size $d/R=5.78$ (medium size).
The vertical lines in Fig.~\ref{fig:TE}(e) represent $d/v$ at each temperature.
We clearly see the transfer entropy peak position is well matched with the estimated time $d/v$, clarifying that the physical meaning of the transfer entropy peak is the information transmission time, or the writing time of information.

\rev{It is notable that the information transmission time is determined not by the diffusion, but by the thermal motion.
The diffusion time is calculated by $\tau_\mathrm{diff} = d^2 / 4\mathcal{D}$, where the diffusion coefficient is given by~\cite{schutte2014inertia}
\begin{equation}
    \mathcal{D} = k_\mathrm{B}T \frac{\alpha D}{(\alpha D)^2 + G^2}.
\end{equation}
The calculated $\tau_\mathrm{diff}$ is two orders of magnitude greater than $d/v$ [While $d/v\sim 10 \, \mathrm{ns}$ as shown in Fig.~\ref{fig:TE}(e), the diffusion time $\tau_\mathrm{diff} \sim 1 \,\mathrm{\mu s}$].
}

Here, we discuss the sensitivity of our results on the number of cells, $N_\mathrm{cell}$.
In the present paper, we have employed $N_\mathrm{cell} = 4$ because of the following two reasons.
First, as $N_\mathrm{cell}$ is increased, the computational complexity of the entropic quantities dramatically increases.
Especially, for the transfer entropy [Eq.~\eqref{eq:TE}], $(N_\mathrm{cell})^3$ combinations of the probability $p(X_t \mid X_{t-\Delta t}, Y_{t-\Delta t})$ must be computed.
\rev{Importantly, the required ensemble size $N$ to suppress statistical error grows rapidly with $N_\mathrm{cell}$.
While it is computationally possible to simulate a higher $N_\mathrm{cell}$, obtaining a sufficient number of ensembles to secure statistical accuracy is often unrealistic in actual experiments.
We deliberately chose the case of $N_\mathrm{cell}=4$, which is experimentally feasible.}
Second, for smaller $N_\mathrm{cell}$ such as $N_\mathrm{cell} = 2$, we cannot analyze the circular motion of the skyrmions.
Importantly, the physical meaning of the transfer entropy, namely ``the state transition time between the stable configurations", cannot be elucidated with $N_\mathrm{cell} < 4$.
Therefore, $N_\mathrm{cell} = 4$ is the most necessary and sufficient division of the confinement box.
If we increase $N_\mathrm{cell}$, the peak position of the transfer entropy would be shifted to a smaller $\Delta t$.
For example, $N_\mathrm{cell} = 16$ ($4 \times 4$ division) would give rise to the peak position of $d/2v$, since the distance between the adjacent cells becomes $d/2$.
%From this argument, we can understand that our conclusion about the transfer entropy is not affected by increasing $N_\mathrm{cell}$.
\rev{From this argument, we can understand that, although the peak position shifts quantitatively, our qualitative conclusion regarding the mechanism of information flow is not affected by increasing $N_\mathrm{cell}$.}

%%%%%%%%%%%%%%%%%%%%%%%%%%%%%%%%%%%%%%%%%%%%%%%%%%%%%%%%%%%%%%%%%%%%%%%%%%%%%%%%%%%%%%%%%%%%%%%%%%%%%%%%%%%%%%%%%%%%%%%%%%%%%%%%
\section{Conclusion}
\label{sec:conclusion}

We theoretically investigated flow of information in a two-skyrmion system confined in a box at a finite temperature, especially focusing on the essential meaning of transfer entropy.
Numerical solutions to the Thiele--Langevin equation indicated that skyrmion dynamics are nontrivial, although this system consists only of two particles.
This complexity arises from the chiral motion of skyrmions resulting from the gyrocoupling term $\bm{G}\times\bm{v}$.
By the coarse-graining of the system using the four cells, the clockwise cyclotron motion with a radius much smaller than the cell size is ignored.
Only the counterclockwise skipping edge motion remains, and as a consequence, the detailed balance condition is violated even in equilibrium, meaning an asymmetric flow of information.

By analyses of this unique system using information-theoretical quantities including Shannon entropy, mutual information, active information storage, and transfer entropy, we elucidated the physical meaning of transfer entropy.
The transfer entropy peak position is independent of the interaction range between the two skyrmions, but is dependent on the box size.
Thus, the transfer entropy peak signifies the time interval required for one particle to influence the other particle’s state, representing the information transmission time.
Contrary to expectations, we showed that the speed of information transfer is unaffected by the interaction strength $\xi$.
Information transmission time can be understood as the sum of the time to obtain mutual information and the time to write the information.
By comparing the estimated characteristic time $d/v$ and the transfer entropy peak position, we further confirmed the above consequence.

%This work provides a essential understanding of transfer entropy and demonstrates the value of information-theoretical analyses in physical systems.
%This study bridges the gap between the conceptual understanding of transfer entropy in complex systems and its application to simpler physical systems, thereby strengthening the connection between information theory and physics.
This work provides an essential understanding of transfer entropy as a measure of flow of information, thereby strengthening the connection between information theory and physics.
Furthermore, these findings could be important to consider a future application such as the ultralow power Brownian computing using the flow of information between skyrmions.
Especially, the violation of the detailed balance condition makes the two-skyrmion system very intriguing \rev{device} for an efficient machine learning algorithm using stochastic systems~\cite{Ohzeki2016stochastic,Takahashi2016,ohzeki2017quantum}.

%%%%%%%%%%%%%%%%%%%%%%%%%%%%%%%%%%%%%%%%%%%%%%%%%%%%%%%%%%%%%%%%%%%%%%%%%%%%%%%%%%%%%%%%%%%%%%%%%%%%%%%%%%%%%%%%%%%%%%%%%%%%%%%%
\begin{acknowledgements}
This work was funded by JSPS KAKENHI Grant Numbers JP23KJ1497, JP20H05666, and JP24K22860, JST CREST Grant Number JPMJCR20C1, and the Spintronics Research Network of Japan (Spin-RNJ).
SM is funded by X-NICS (JPJ011438).
\end{acknowledgements}

\bibliography{reference}
%%%%%%%%%%%%%%%%%%%%%%%%%%%%%%%%%%%%%%%%%%%%%%%%%%%%%%%%%%%%%%%%%%%%%%%%%%%%%%%%%%%%%%%%%%%%%%%%%%%%%%%%%%%%%%%%%%%%%%%%%%%%%%%%

%%%%%%%%%%%%%%%%%%%%%%%%%%%%%%%%%%%%%%%%%%%%%%%%%%%%%%%%%%%%%%%%%%%%%%%%%%%%%%%%%%%%%%%%%%%%%%%%%%%%%%%%%%%%%%%%%%%%%%%%%%%%%%%%
\clearpage
\onecolumngrid

\begin{center}
    \textbf{\large Supplemental Material for ``Transfer Entropy and Flow of Information in Two-Skyrmion System"}
\end{center}

% --- カウンターのリセット ---
\setcounter{equation}{0}
\setcounter{figure}{0}
\setcounter{table}{0}
\setcounter{page}{1} % ページ番号も1からリセットしたい場合
\setcounter{section}{0} % セクション番号もリセットする場合
% --- 表示形式の変更 (S1, S2... の形式にする) ---
\renewcommand{\theequation}{S\arabic{equation}}
\renewcommand{\thefigure}{S\arabic{figure}}
\renewcommand{\thetable}{S\arabic{table}}
\renewcommand{\thesection}{S\arabic{section}}

\maketitle

%%%%%%%%%%%%%%%%%%%%%%%%%%%%%%%%%%%%%%%%%%%%%%%%%%%%%%%%%%%%%%%%%%%%%%%%%%%%%%%%%%
\section{Analytical solution of master equation}
\label{app:master}
In this appendix, we present details of the analysis using the master equation shown in Figs.~\ref{fig:probability} and \ref{fig:prob2body}.
The classical master equation is written as
\begin{gather}
    \frac{d\bm{q}_t}{dt} = A \bm{q}_t, \nonumber
    \\
    A = \begin{pmatrix} -(w+w'+w'') & w' & w'' & w \\ w & -(w+w'+w'') & w' & w'' \\ w'' & w & -(w+w'+w'') & w' \\ w' & w'' & w & -(w+w'+w'') \end{pmatrix},
    \\
    \bm{q}_t = \begin{pmatrix} q_t(0) \\ q_t(1) \\ q_t(2) \\ q_t(3) \end{pmatrix}, \nonumber
\end{gather}
where $q_t(x)$ is the probability that a skyrmion exists in cell $x$ at certain time $t$ ($x = 0, 1, 2, 3$).
The parameters $w,w'$, and $w''$ represents the transition rate between the cells $0\rightarrow1$, $0\rightarrow3$, and $0\rightarrow2$, respectively.
When the smallest parameter $w''$ was turned off, we were able to solve the equation analytically:
\begin{equation}
    \begin{pmatrix} q_t(0) \\ q_t(1) \\ q_t(2) \\ q_t(3) \end{pmatrix} = \frac{1}{4} \left[\begin{pmatrix} 1 \\ 1 \\ 1 \\ 1 \end{pmatrix} + \begin{pmatrix} 1 \\ -1 \\ 1 \\ -1 \end{pmatrix}e^{-2w_{+}t} + \begin{pmatrix} \cos{w_{-}t} \\ \sin{w_{-}t} \\ -\cos{w_{-}t} \\ -\sin{w_{-}t} \end{pmatrix}e^{-w_{+}t}  \right],
\end{equation}
where $w_{\pm} = w \pm w'$.
We set the initial condition $\bm{q}_{t=0} = (1,0,0,0)^\top$.
In Figs.~\ref{fig:probability} and \ref{fig:prob2body}, we fitted the simulated results using this analytical form.
Note that the inclusion of $w''$ does not largely alter the result.

Regarding the detailed balance condition, we calculate Eq.~\eqref{eq:detail}.
At equilibrium, we have
\begin{align}
\begin{split}
    r_{0\rightarrow1}(\Delta t) &= 
    \frac{p(x_t=1 \mid x_{t-\Delta t}=0)}{p(x_t=0 \mid x_{t-\Delta t}=1)}
    \\
    &= 
    \frac{p(x_t=1 \mid x_{t-\Delta t}=0)}{p(x_t=3 \mid x_{t-\Delta t}=0)}
    \\
    &=
    \frac{q_{\Delta t}(1)}{q_{\Delta t}(3)}
    \\
    &=
    \frac{1 - e^{-2w_+ \Delta t} + \sin{(w_- \Delta t)}e^{-w_+ \Delta t}}
    {1 - e^{-2w_+ \Delta t} - \sin{(w_- \Delta t)}e^{-w_+ \Delta t}},
\end{split}
\label{eq:detail-master}
\end{align}
since $p(x_{t-\Delta t}=0) = p(x_{t-\Delta t}=1) = 1/4$ and the system has the 4-fold rotational symmetry.
From Eq.~\eqref{eq:detail-master}, we find $r_{0 \rightarrow 1}$ deviates from unity if $w_- = w - w'$ is finite.
Therefore, the detailed balance condition is violated at the level of the master equation.
%%%%%%%%%%%%%%%%%%%%%%%%%%%%%%%%%%%%%%%%%%%%%%%%%%%%%%%%%%%%%%%%%%%%%%%%%%%%%%%%%%

%%%%%%%%%%%%%%%%%%%%%%%%%%%%%%%%%%%%%%%%%%%%%%%%%%%%%%%%%%%%%%%%%%%%%%%%%%%%%%%%%%
\section{Analytical formula for energetic box}
\label{app:corner}

\begin{figure*}
\begin{center}
   \includegraphics [width=0.9\linewidth]{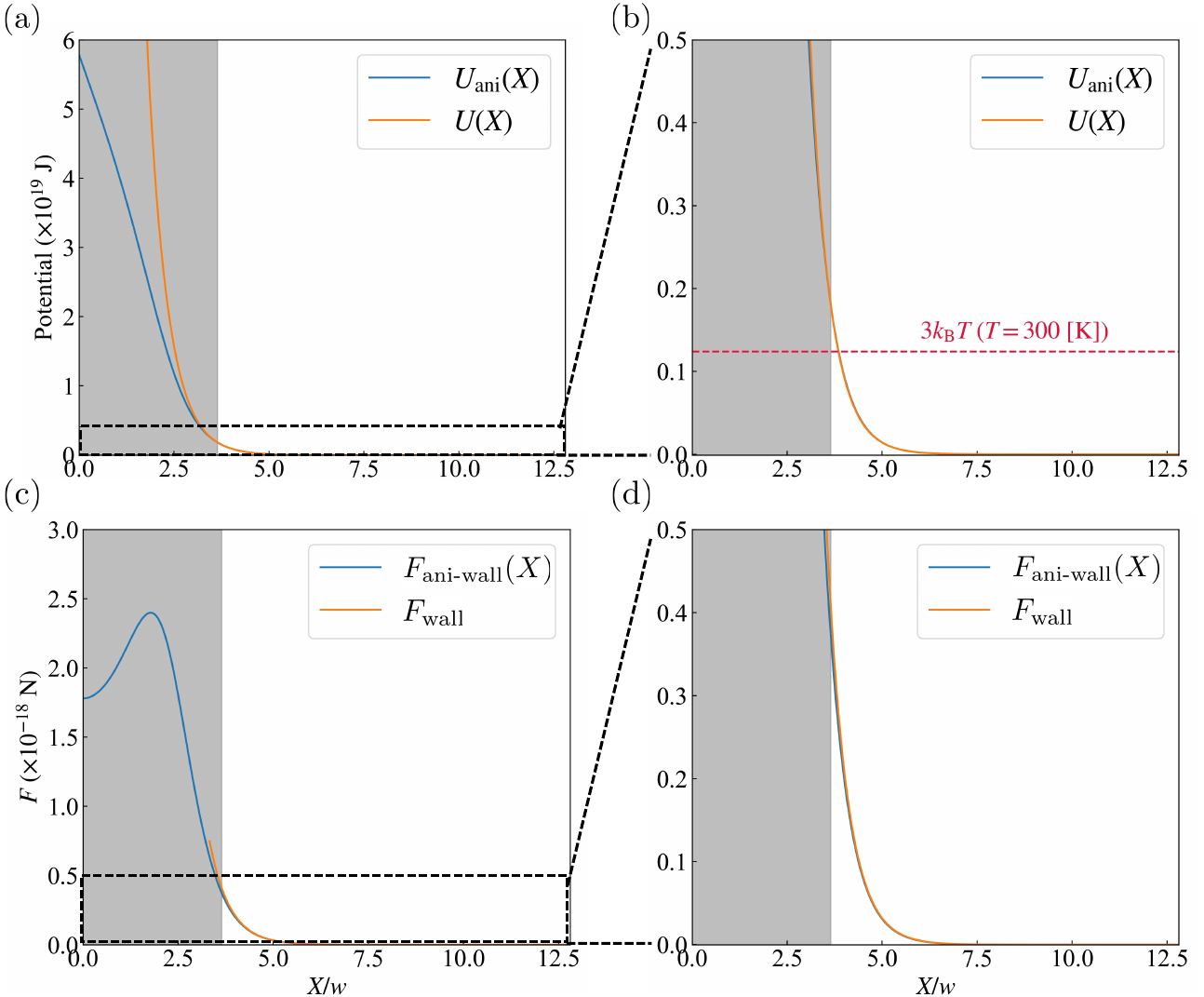}
   \caption{ 
   (a), (b) Skyrmion potential formed by the magnetic anisotropy barrier and (c), (d) forces acting on the skyrmion.
   The blue and orange curves show the analytical equation and the model equation, respectively.
   (b) and (d) are enlarged views of (a) and (b), respectively.
   The gray areas in (b) and (d) are not considered in the simulations because of the large contribution from the corners.
            }\label{fig:anicompare}
 \end{center}
 \end{figure*}

\begin{figure*}
\begin{center}
   \includegraphics [width=\linewidth]{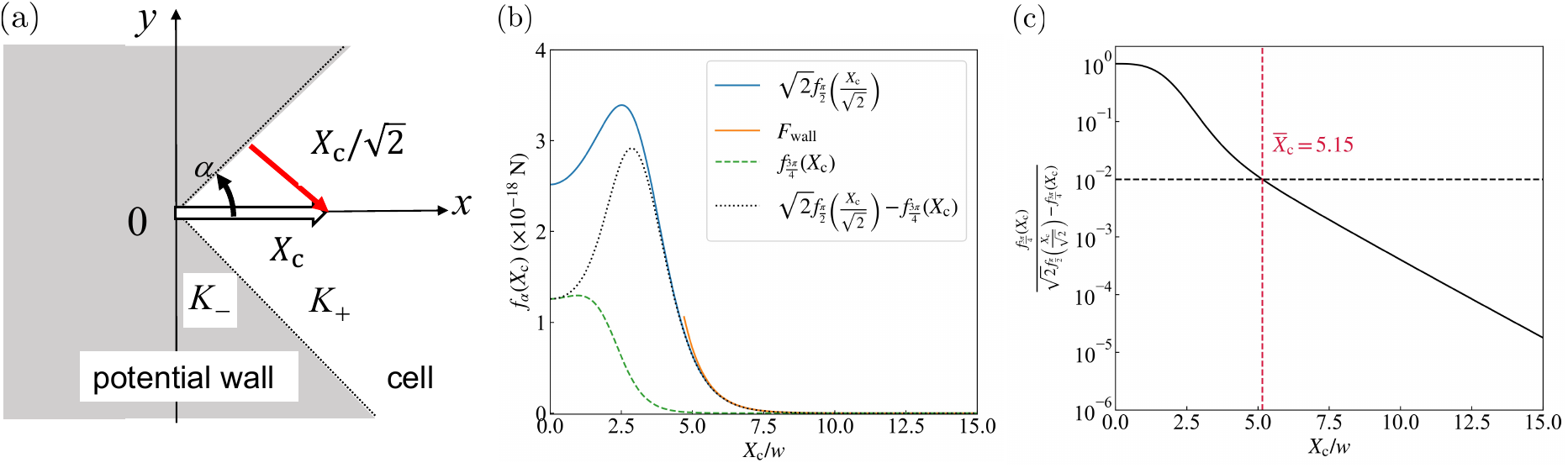}
   \caption{ 
   Evaluation of forces acting from corner walls.
   (a) Coordinate system of the corner.
   Gray: magnetic anisotropic barrier.
   White: cell interior.
   (b) Forces acting from the wall. The blue and orange curves are the force $f_{\pi/2}$ from the wall and its model function $F_\mathrm{wall}$ at the corner angle $\alpha=\pi/2$, respectively.
   The green dashed and black dotted lines are the forces $f_{3\pi/4}$ and the difference between $f_{\pi/2}$ and $f_{3\pi/4}$.
   (c) Ratio of $f_{\pi/2}$ to $\sqrt{2}f_{\pi/2} - f_{3\pi/4}$.
   The ratio of forces at $X_\mathrm{c}/w = 5.15$ is $10^{-2}$.
            }\label{fig:corner}
 \end{center}
 \end{figure*}

In this section, we evaluate the validity of modeling the force from the potential barrier to the skyrmion.
In this study, we considered barriers that confine skyrmion using a confinement-efficient magnetic anisotropy energy step~\cite{miki2021size,tamura2020skyrmion}.
The potential $U_{\mathrm{ani}} (X)$ of a skyrmion at a distance of $X$ from the wall owing to the magnetic anisotropy energy step and the force $\bm{F}_{\mathrm{ani\text{-}wall}}$ from the energy barrier obtained from this potential $U_{\mathrm{ani}} (X)$ are respectively described in the following equations~\cite{tamura2020skyrmion}
\begin{align}
\begin{split}
    U_{\mathrm{ani}}(X) &= 
    -\Delta Kt \int_{-\infty}^{\infty} \int_{0}^{X}dxdy (1-m_z^2 (\bm{r})) 
    + \frac{K_\mathrm{wall} + K_\mathrm{cell}}{2} t \int_{-\infty}^{\infty} \int_{-\infty}^{\infty}dxdy (1-m_z^2 (\bm{r}))
    \\
    &= -2\Delta Kt \int_{0}^{\infty} \int_{0}^{X}dxdy
    \frac{4 \sinh^2{(R/w)} \sinh^2{(\sqrt{x^2+y^2}/w)} }
    {[\sinh^2{(R/w)} + \sinh^2{(\sqrt{x^2+y^2}/w)}]^2}
    \\
    & \quad +
    2(K_\mathrm{wall} + K_\mathrm{cell})t
    \int_{0}^{\infty} \int_{0}^{\infty}dxdy
    \frac{4 \sinh^2{(R/w)} \sinh^2{(\sqrt{x^2+y^2}/w)} }
    {[\sinh^2{(R/w)} + \sinh^2{(\sqrt{x^2+y^2}/w)}]^2},
\end{split}
\label{eq:Uani}
\end{align}
\begin{equation}
    F_{\mathrm{ani\text{-}wall}}(X) = -\Delta Kt \int_{-\infty}^{\infty}dy
    \frac{4 \sinh^2{(R/w)} \sinh^2{(\sqrt{x^2+y^2}/w)} }
    {[\sinh^2{(R/w)} + \sinh^2{(\sqrt{x^2+y^2}/w)}]^2},
\label{eq:Faniwall}
\end{equation}
where $t$ and $m_z(\bm{r})$ are the thickness of the ferromagnetic film and the $z$-component of the magnetization at position $\bm{r} = (x,y) = (r\cos{\varphi}, r\sin{\varphi})$, respectively.
$K_\mathrm{wall}$ and $K_\mathrm{cell}$ are the magnetic anisotropy at the potential barrier and the cell, respectively.
$\Delta K = K_\mathrm{wall} - K_\mathrm{cell}$ is the height of the magnetic anisotropy energy step.
The skyrmion profiles are described by the functions modeled in Ref.~\cite{wang2018theory}.
The second term of $U_\mathrm{ani}(X)$ is a constant, and therefore does not contribute to the force $\bm{F}_{\mathrm{ani\text{-}wall}}$.
Because this function is complex and requires an enormous amount of time to simulate, an exponential model function [Eq.~\eqref{eq:Uwall}] was used.
Figure~\ref{fig:anicompare}(a) shows the potential $U_{\mathrm{ani}}(X)$~[Eq.~\eqref{eq:Uani}, blue curve] and the model function $U_\mathrm{wall}$~[Eq.~\eqref{eq:Uwall}, orange curve].
Fig.~\ref{fig:anicompare}(c) shows the force $F_{\mathrm{ani\text{-}wall}}$ from the wall~[Eq.~\eqref{eq:Faniwall}, blue curve] and the model function $F_\mathrm{wall}$ [Eq.~\eqref{eq:Fwall}, orange curve].
Figure~\ref{fig:anicompare}(b) and (d) are the enlarged views of (a) and (c), respectively.
The potential $U_\mathrm{ani}(X)$ and force $F_{\mathrm{ani\text{-}wall}}$ from the wall shown in Fig.~\ref{fig:anicompare} correspond to $K_\mathrm{cell} = 0.9 \, \mathrm{MJ/m^3}$, $K_\mathrm{wall} = 1.15K_\mathrm{cell} = 1.035 \, \mathrm{MJ/m^3}$, and $t=1.2 \, \mathrm{nm}$.
In Fig.~\ref{fig:anicompare}(b), the red dashed line represents the value of $3k_\mathrm{B}T$ at $300 \, \mathrm{K}$, which means that the skyrmions are confined within a standard deviation of $3\sigma$ in the cell.
In the region $X>1.61R$, the model functions for the potential and force were fitted to the analytical equations $U_\mathrm{ani}(X)$ and $F_{\mathrm{ani\text{-}wall}}$.
These model functions only consider the direction perpendicular to the wall and do not include the contribution from the wall at the corners of the cell. 
Therefore, the effects of the corners are evaluated below.
Fig.~\ref{fig:corner}(a) shows the coordinate system of the cell corner rotated by $\pi/4$ from the coordinate system in the main text.
The gray area represents the potential barrier, while the white area represents the cell interior.
In general, the formula for a potential with a corner of angle $\alpha$ is
\begin{align}
\begin{split}
    U_\mathrm{ani}(X) &= -2\Delta Kt \int_0^\alpha d\varphi \int_0^\infty rdr (1-m_z^2 (\bm{r}))
    + 2K_\mathrm{wall}t \int_0^\pi d\varphi \int_0^\infty rdr (1-m_z^2 (\bm{r}))
    \\
    &= -2\Delta Kt \int_0^\alpha d\varphi \int_0^\infty rdr
    \frac{4 \sinh^2{(R/w)} \sinh^2{(\sqrt{r^2+X^2-2rX\cos{\varphi}}/w)} }
    {[\sinh^2{(R/w)} + \sinh^2{(\sqrt{r^2+X^2-2rX\cos{\varphi}}/w)}]^2}
    \\
    & \quad +
    2K_\mathrm{wall}t \int_0^\pi d\varphi \int_0^\infty rdr
    \frac{4 \sinh^2{(R/w)} \sinh^2{(\sqrt{r^2+X^2-2rX\cos{\varphi}}/w)} }
    {[\sinh^2{(R/w)} + \sinh^2{(\sqrt{r^2+X^2-2rX\cos{\varphi}}/w)}]^2}.
\end{split}
\end{align}
Furthermore, the force $f_\alpha$ acting on the corner is
\begin{align}
\begin{split}
    & f_\alpha(X)
    = -\Delta Kt \int_0^\alpha d\varphi \int_0^\infty rdr \frac{1}{w}
    \frac{X-r\cos{\varphi}}
    {\sqrt{r^2+X^2-2rX\cos{\varphi}}}
    \\
    & \times 
    \frac{4 \sinh^2{(R/w)} \sinh{(2\sqrt{r^2+X^2-2rX\cos{\varphi}}/w)} }
    {[\sinh^2{(R/w)} + \sinh^2{(\sqrt{r^2+X^2-2rX\cos{\varphi}}/w)}]^2}
    \\
    & \times 
    \left[1 - 
    \frac{2\sinh^2{(\sqrt{r^2+X^2-2rX\cos{\varphi}}/w)}}
    {\sinh^2{(R/w)} + \sinh^2{(\sqrt{r^2+X^2-2rX\cos{\varphi}}/w)}}
    \right].
\end{split}
\end{align}
Figure~\ref{fig:corner}(b) shows the dependence of distance from the corner $X_\mathrm{c}$ to the force in the coordinate system in (a).
The blue and green solid curves represent the forces $f_{\pi/2}$ and $f_{3\pi/4}$, respectively, and the black dotted curve represents the difference between $f_{\pi/2}$ and $f_{3\pi/4}$.
The orange curve represents the model function $F_{\mathrm{wall}}$.
$f_{\pi/2}$ and $F_{\mathrm{wall}}$ take the factor $\sqrt{2}$, because the coordinates were transformed from Eq.~\eqref{eq:Fwall} by angle $\pi/4$.
Figure~\ref{fig:corner}(c) is the ratio of $f_{\pi/2}$ to $\sqrt{2}f_{\pi/2} - f_{3\pi/4}$, which was used to evaluate the contribution of the corner barrier.
The force ratio is less than $10^{-2}$ in the region $X_\mathrm{c}/w > 5.15$, i.e., $X>1.61R$, and the corner components were very small, except for the gray regions in Fig.~\ref{fig:anicompare}(b) and (d).
Therefore, to reduce the simulation time, we evaluated various flows of information in the region $X>1.61R$ using a simple model function in Eqs.~\eqref{eq:Fwall} and \eqref{eq:Uwall}, respectively.
 %%%%%%%%%%%%%%%%%%%%%%%%%%%%%%%%%%%%%%%%%%%%%%%%%%%%%%%%%%%%%%%%%%%%%%%%%%%%%%%%%%
 \section{Correlation function}
\label{app:corr}
\begin{figure}
\begin{center}
   \includegraphics [width=0.5\linewidth]{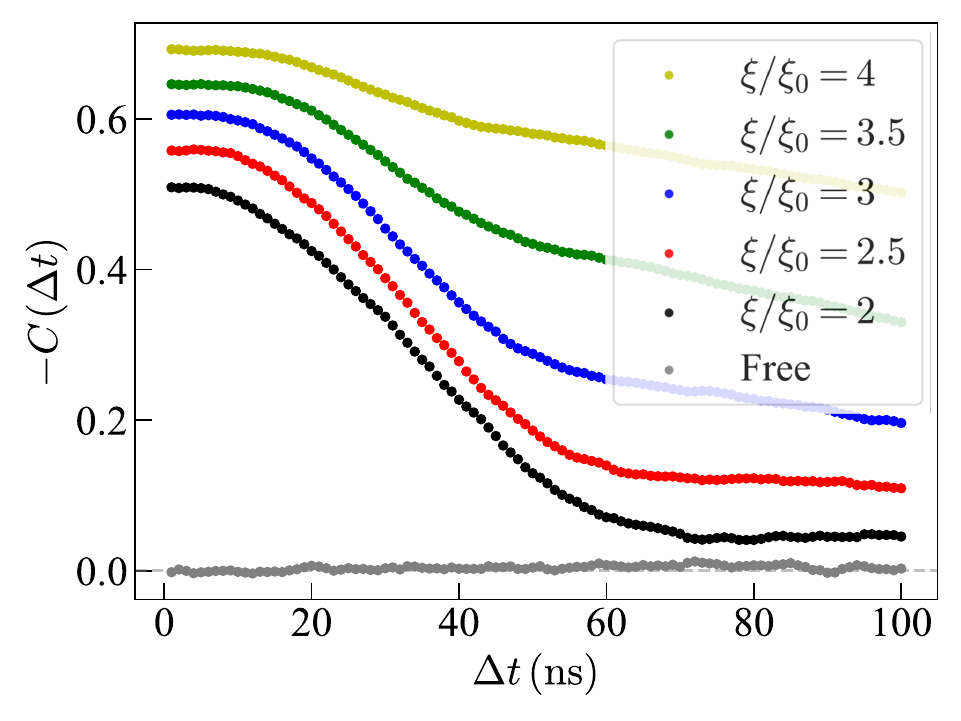}
   \caption{ 
   Plot of $-C(\Delta t)$ (the correlation function $C$ with the sign reversed) as a function of the time delay $\Delta t$.
   The different colors correspond to the different interaction ranges $\xi$ as illustrated by the legend.
            }\label{fig:corr-func}
 \end{center}
 \end{figure}

Here, we calculate the correlation function between the skyrmion $X$ and $Y$.
First, we focus on the states of the two-skyrmion system in terms of the deviation from the expectation value,
\begin{equation}
\begin{split}
    & \delta x_t^{(i)} = x_t^{(i)} - \langle x_t \rangle,
    \\
    & \delta y_t^{(i)} = y_t^{(i)} - \langle y_t \rangle,
\end{split}
\end{equation}
at $t = t_\mathrm{f}$.
The expectation value $\langle x_t \rangle = \langle y_t \rangle = 1.5$.
The correlation function is then defined by
\begin{equation}
    C(\Delta t) = \frac{1}{N} \sum_{i=1}^{N} \delta x_t^{(i)} \delta y_{t-\Delta t}^{(i)}.
\end{equation}
For sufficiently large time delay $\Delta t$, the correlation function $C(\Delta t)$ vanishes.
Since the stable configurations $(x_t, y_t) = (0,2), (1,3)$ correspond to $(\delta x_t, \delta y_t) = (-1.5, +0.5), (-0.5, +1.5)$, $C(\Delta t)$ takes negative values when the two skyrmions are correlated.
Figure~\ref{fig:corr-func} shows $-C(\Delta t)$ as a function of $\Delta t$.
$-C(\Delta t)$ monotonically decreases in the similar manner as the mutual information [see Fig.~\ref{fig:ShannonMI}(b)], and importantly, it exhibits no peak structure.
This result indicates that the analysis using the correlation function is not sufficient to detect the information transmission time, and the transfer entropy is necessary to study the flow of information.

%%%%%%%%%%%%%%%%%%%%%%%%%%%%%%%%%%%%%%%%%%%%%%%%%%%%%%%%%%%%%%%%%%%%%%%%%%%%%%%%%%

%%%%%%%%%%%%%%%%%%%%%%%%%%%%%%%%%%%%%%%%%%%%%%%%%%%%%%%%%%%%%%%%%%%%%%%%%%%%%%%%%%
\section{Details of numerical simulations}
\label{app:numerical}

\rev{
In this study, the Thiele--Langevin equation [Eq.~\eqref{eq:TL}] was numerically solved using the fourth-order Runge--Kutta method as follows.
The states of the two skyrmions ($j=1,2$) are updated by
\begin{equation}
    \begin{split}
    & v_{jx} (t+\delta t) = v_{jx}(t) + \varphi(f_j(t,\bm{\Gamma}(t),\zeta_{jx})) \delta t, 
    \\
    & v_{jy} (t+\delta t) = v_{jy}(t) + \varphi(g_j(t,\bm{\Gamma}(t),\zeta_{jy})) \delta t, 
    \\
    & x_j(t+\delta t) = x_j(t) + \varphi(v_{jx}(t)) \delta t,
    \\
    & y_j(t+\delta t) = y_j(t) + \varphi(v_{jy}(t)) \delta t,
    \end{split}
\end{equation}
where $\zeta_{jx},\zeta_{jy} \sim \mathcal{N}(0,1)$ are stochastic variables, $\bm{\Gamma} = (\bm{r}_1,\bm{r}_2,\bm{v}_1,\bm{v}_2)$, and
\begin{equation}
    \begin{split}
        & f_j(t,\bm{\Gamma}(t),\zeta_{jx}) 
        = \frac{1}{m} 
        \left[-\alpha D v_x + Gv_y + (-1)^{j-1}F_\mathrm{int}^x + F_\mathrm{wall}^x +  \sqrt{\frac{2\alpha D k_\mathrm{B}T}{\delta t}} \zeta_{jx} \right],
        \\
        & g_j(t,\bm{\Gamma}(t),\zeta_{jy}) 
        =\frac{1}{m} \left[-\alpha D v_y - Gv_x + (-1)^{j-1}F_\mathrm{int}^y + F_\mathrm{wall}^y + \sqrt{\frac{2\alpha D k_\mathrm{B}T}{\delta t}} \zeta_{jy} \right].
    \end{split}
\end{equation}
Here, $\bm{F}_\mathrm{int} = (F_\mathrm{int}^x, F_\mathrm{int}^y)$ and $\bm{F}_\mathrm{wall} = (F_\mathrm{wall}^x, F_\mathrm{wall}^y)$.
The function $\varphi$ is defined as
\begin{equation}
    \begin{split}
    & \varphi(f(t,\bm{\Gamma}(t),\zeta)) = \frac{k_1+2k_2+2k_3+k_4}{6},
    \\
    & k_1 = f(t,\bm{\Gamma}(t),\zeta),
    \\
    & k_2 = f(t+\delta t/2,\bm{\Gamma}(t)+k_1\delta t/2,\zeta),
    \\
    & k_3 = f(t+\delta t/2,\bm{\Gamma}(t)+k_2\delta t/2,\zeta),
    \\
    & k_4 = f(t+\delta t,\bm{\Gamma}(t)+k_3\delta t,\zeta).
    \end{split}
\end{equation}
}
%%%%%%%%%%%%%%%%%%%%%%%%%%%%%%%%%%%%%%%%%%%%%%%%%%%%%%%%%%%%%%%%%%%%%%%%%%%%%%%%%%

%%%%%%%%%%%%%%%%%%%%%%%%%%%%%%%%%%%%%%%%%%%%%%%%%%%%%%%%%%%%%%%%%%%%%%%%%%%%%%%%%%
\section{Skyrmion profile and dissipation dyadic}
\revblue{
For the calculation of dissipation dyadic, we use Eq.~\eqref{eq:D-mag},
\begin{equation}
\label{eq:D-supple}
    D = \frac{M_\mathrm{s}h}{|\gamma_e|} \int d^2\bm{r} \,\frac{\partial\bm{m}}{\partial x} \cdot \frac{\partial\bm{m}}{\partial x},
\end{equation}
where $\bm{m}(\bm{r}) = \bm{M}(\bm{r}) / M_\mathrm{s}$.
In general, dissipation dyadic is a tensor, $D_{ij} \, (i,j=x,y)$.
However, in this paper, we consider a rigid-body skyrmion and therefore the axial symmetry is preserved, giving $D\equiv D_{xx}=D_{yy}$ and $D_{xy}=D_{yx}=0$.
The normalized magnetization is expressed as
\begin{equation}
    \bm{m}(\bm{r}) = 
    \begin{pmatrix}
        \sin{\Theta(\bm{r})}\cos{\Phi(\bm{r})} \\ \sin{\Theta(\bm{r})}\sin{\Phi(\bm{r})} \\ \cos{\Theta(\bm{r})}
    \end{pmatrix}
    ,
\end{equation}
where $\Phi$ and $\Theta$ are the azimuthal and polar angles of $\bm{m}$, respectively.
As the skyrmion's magnetization profile, we use a $360^\circ$ domain-wall model~\cite{wang2018theory},
\begin{equation}
\begin{split}
    & \Theta(\bm{r}) = 2\arctan{ \left[ \frac{ \sinh{(r/w)} }{ \sinh{(R/w)} } \right]},
    \\
    & \Phi(\bm{r}) = q\varphi + \gamma.
\end{split}
\end{equation}
Here, we employ polar coordinates $\bm{r} = (r,\varphi)$; the skyrmion is centered at $r=0$.
We use the skyrmion number $q=+1$.
The symbol $\gamma$ denotes the helicity of skyrmion.
We consider N\'{e}el skyrmion, which has $\gamma = 0$ or $\pi$.
This model for the skyrmion's profile and Eq.~\eqref{eq:D-supple} reads~\cite{wang2018theory}
\begin{align}
    &D = \frac{2\pi M_\mathrm{s}h}{|\gamma_e|} f(R/w),
    \\
    &f(x) = \int_0^\infty dt\, \left[ 
    \frac{2\sinh^2{x}\cosh^2{t}}{(\sinh^2{x}+\sinh^2{t})^2}t + \frac{2\sinh^2{x}\sinh^2{t}}{t(\sinh^2{x}+\sinh^2{t})^2}
    \right].
\end{align}
By numerically calculating $f(R/w)$ with parameters $R=25.55 \, \mathrm{nm}$ and $w=11.35 \, \mathrm{nm}$~\cite{tamura2020skyrmion,cho2020manipulating,wang2018theory,hrabec2017current,belavin1975metastable}, we obtain $D=7.06\times10^{-14} \, \mathrm{kg/s}$.
}

%%%%%%%%%%%%%%%%%%%%%%%%%%%%%%%%%%%%%%%%%%%%%%%%%%%%%%%%%%%%%%%%%%%%%%%%%%%%%%%%%%

%%%%%%%%%%%%%%%%%%%%%%%%%%%%%%%%%%%%%%%%%%%%%%%%%%%%%%%%%%%%%%%%%%%%%%%%%%%%%%%%%%%%%%%%%%%%%%%%%%%%%%%%%%%%%%%%%%%%%%%%%%%%%%%%
\end{document}